\documentclass[fleqn,usenatbib]{mnras}

\usepackage{newtxtext,newtxmath}

\usepackage[T1]{fontenc}

\DeclareRobustCommand{\VAN}[3]{#2}
\let\VANthebibliography\thebibliography
\def\thebibliography{\DeclareRobustCommand{\VAN}[3]{##3}\VANthebibliography}

\usepackage{graphicx}	
\usepackage{amsmath}	
\usepackage{threeparttable}
\usepackage{subfig,float}
\usepackage{comment}
\usepackage{appendix}
\usepackage{ragged2e}
\usepackage{lineno}
\usepackage[normalem]{ulem}

\newcommand{\hii}{H{\sc ii}}

\title[SMGPS filamentary source catalogue]{The SARAO MeerKAT Galactic Plane Survey filamentary source catalogue}

\author[G. M. Williams et al.]{G. M. Williams,$^{1,2}$\thanks{E-mail: g.williams@aber.ac.uk (GMW)}
M. A. Thompson,$^{2}$\thanks{Email: M.A.Thompson@leeds.ac.uk (MAT)}
M. Mutale,$^{2}$
A. J. Rigby,$^{2}$
C. Bordiu,$^{3}$
S. Riggi,$^{3}$
\newauthor
M. Bietenholz,$^{4,5}$
L. D. Anderson,$^{6,7,8}$
F. Camilo,$^{9}$
S. Goedhart,$^{9,10}$
S. E. Jaffa,$^{11}$
W. O. Obonyo,$^{12,13}$
\newauthor
C. Trigilio,$^{3}$
and G. Umana$^{3}$
\\
$^{1}$Department of Physics, Aberystwyth University, Ceredigion, Cymru, SY23 3BZ, UK\\
$^{2}$School of Physics \& Astronomy, University of Leeds, Leeds, LS2 9JT, UK\\
$^{3}$INAF-Osservatorio Astrofisico di Catania, Via Santa Sofia 78, 95123 Catania, Italy\\
$^{4}$SARAO/Hartebeesthoek Radio Astronomy Observatory, PO Box 443, Krugersdorp 1740, South Africa\\
$^{5}$Department of Physics and Astronomy, York University, Toronto, M3J 1P3, Ontario, Canada\\
$^{6}$Department of Physics and Astronomy, West Virginia University, Morgantown, WV 26506, USA\\
$^{7}$Adjunct Astronomer at the Green Bank Observatory, P.O. Box 2, Green Bank, WV 24944, USA\\
$^{8}$Center for Gravitational Waves and Cosmology, West Virginia University, Chestnut Ridge Research Building, Morgantown, WV 26505, USA\\
$^{9}$South African Radio Astronomical Observatory, 2 Fir Street, Black River Park, Observatory, Cape Town, 7925, South Africa\\
$^{10}$SKA Observatory, 2 Fir Street, Observatory 7925, Cape Town, South Africa\\
$^{11}$Research IT, University of Manchester, Oxford Road, Manchester, M13 9PL, UK\\
$^{12}$Department of Mathematical Sciences, University of South Africa, Cnr Christian de Wet Rd and Pioneer Avenue, Florida Park, 1709, Roodepoort, South Africa\\
$^{13}$Department of Astronomy and Space Science, The Technical University of Kenya, P.O. Box 52428 - 00200, Nairobi, Kenya\\
}

\date{Accepted XXX. Received YYY; in original form ZZZ}

\pubyear{2023}

\begin{document}
\label{firstpage}
\pagerange{\pageref{firstpage}--\pageref{lastpage}}
\maketitle

\begin{abstract}
We present a catalogue of filamentary structures identified in the SARAO (South African Radio Astronomy Observatory) MeerKAT 1.3\,GHz Galactic Plane Survey (SMGPS).  We extract 933 filaments across the survey area, 803 of which (${\sim}86\%$) are associated with extended radio structures (e.g. supernova remnants and \hii\ regions), whilst 130 (${\sim}14\%$) are largely isolated. We classify filaments as thermal or non-thermal via their associated mid-infrared emission and find 77/130 (${\sim}59\%$) of the isolated sources are likely to be non-thermal, and are therefore excellent candidates for the first isolated, non-thermal radio filaments observed outside of the Galactic Centre (GC). Comparing the morphological properties of these non-thermal candidates to the non-thermal filaments observed towards the GC we find the GC filaments are on the whole angularly narrower and shorter than those across the SMGPS, potentially an effect of distance. The SMGPS filaments have flux densities similar to those of the GC, however the distribution of the latter extends to higher flux densities. If the SMGPS filaments were closer than the GC population, it would imply a more energetic population of cosmic ray electrons in the GC. We find the filament position angles in the SMGPS are uniformly distributed, implying that the local magnetic field traced by the filaments does not follow the large-scale Galactic field. Finally, although we have clearly shown that filaments are not unique to the GC, the GC nevertheless has the highest density of filaments in the Milky Way. 
\end{abstract}

\begin{keywords}
catalogues -- surveys -- techniques: interferometric -- Galaxy: structure -- ISM: structure -- radio continuum: ISM
\end{keywords}




\section{Introduction}
\label{sec:intro}

Our understanding of the structure of the Milky Way has taken great strides in recent years in part due to large surveys of our Galactic Plane at a range of wavelengths. Structures of a filamentary morphology permeate the interstellar medium (ISM) across different size scales, environments and tracers, from low-density dust cirrus in the infrared \citep[][]{Low1984,Jackson2003,Miville-Deschenes2010,Bianchi2017}, to neutral atomic H{\sc i} filaments \citep[e.g.][]{Kalberla16,Kalberla2020,Verschuur2018,Clark2019,Soler2020}, and high-extinction star-forming filaments of dust and molecular gas with a range of scales in the infrared and submillimetre \citep[e.g.][]{Schneider1979,Andre14,Ragan14,Li2016,Mattern18,Zucker2018,Arzoumanian19,Schisano2020}. 
In the radio, tracing the ionised ISM, mysterious filamentary structures have been identified towards the Galactic Centre \citep[GC;][]{Yusef-Zadeh1984,Yusef-Zadeh2004,LawYZC2008,MorrisZG2017,Barkov2019}. 
With observations at 20\,cm (${\sim}$1.5\,GHz) from the Very Large Array (VLA), \cite{Yusef-Zadeh2004} identified more than 80 bright and highly linear filaments (from sub-arcminute to almost half a degree in length). 
Striking images of these radio filaments were recently taken by MeerKAT at 1.28\,GHz \citep{MeerKATCollaboration2018,Heywood2019,Heywood2022}. The unprecedented sensitivity and $u,v$-coverage of MeerKAT allowed for one of the most comprehensive studies of the GC yet, revealing for the first time a population of fainter and shorter filaments \citep{Yusef-Zadeh+2022A,Yusef-Zadeh+2023}. Many of these filaments have non-thermal spectral indices \citep{Yusef-Zadeh+2022A}, and therefore are often referred to as non-thermal radio filaments (NRFs).

Although these NRFs were first discovered 40 years ago, a consensus is yet to be reached on their nature and origin. Given their non-thermal emission, they are reasonably supposed to be synchrotron emission from either an external or in-situ acceleration mechanism of relativistic particles \citep[e.g.][]{Barkov2019, Yusef-ZadehW2019}.
Some postulate that NRFs are magnetic flux tubes illuminated by relativistic particles \citep{Yusef-Zadeh1984,MorrisS1996} from ram pressure-confined pulsar wind nebulae \citep[PWNe; e.g.][]{Bykov+2017,ThomasPE2020}, supernova remnants \citep[SNRs; e.g.][]{Barkov2019}, or stellar winds from massive stars \citep{RosnerB1996}. Many NRFs are indeed seen to terminate around radio sources \citep{Yusef-Zadeh+2022C}. 
The destruction of molecular clouds by the gravitational potential of the GC has also been shown to produce elongated structures with conditions conducive for particle acceleration through magnetic re-connection \citep{CoughlinNG2021}. 
Recent observations with MeerKAT show spatial coincidence of GC NRFs with a large scale radio bubble, the remnant of a past energetic explosion, suggesting they are brightened at the bubble edge \citep{Heywood2022}. Radio observations of galaxy clusters have also identified NRFs in the intracluster medium \cite[ICM; e.g.][]{Ramatsoku+2020,Knowles+2022}. Comparing the ICM and GC NRFs, \cite{Yusef-Zadeh+2022D} suggest both populations may have a common origin either due to compression or stretching of magnetic field lines due to turbulence, or cloud-wind interactions. 
Searches for NRFs in the wider Galactic Plane have so far been negative \citep{Yusef-Zadeh2004}, with the strong implication that to form NRFs in the Milky Way requires the unique high cosmic ray environment of the GC.

The high fidelity and sensitivity of the recent SARAO (South African Radio Astronomy Observatory) MeerKAT 1.3\,GHz Galactic Plane Survey \citep[SMGPS;][]{Goedhart+2024} has revealed a population of filamentary structures across the Galactic Plane. Some of these filaments are convincing candidates for NRFs and, if so, would be the first such observed in the Galactic Plane. 
Confirming these candidates is important for a number of reasons. Firstly, confirming them would immediately call into doubt models for the NRFs that are unique to the peculiar properties of the Galactic Centre \citep[e.g.][]{Yusef-ZadehW2019}. Secondly, the NRFs can act as tracers of relativistic particles injected into the ISM --- tracing the spatial distribution and populations of Galactic cosmic ray sources for the first time \citep[e.g.][]{Zweibel2017}. And lastly, if distances to the NRFs can be assigned they can also act as local measures of the magnetic field strength within the Milky Way, assuming equipartition and that they trace magnetic flux tubes.

In this paper, we present a full catalogue of filaments identified within the SMGPS, including candidate NRFs. In Section~\ref{sec:obs}, we summarise the pertinent details of the SMGPS observations. Our source extraction method is presented in Section~\ref{sec:method}, and the derivation of filament positional and morphological properties is discussed in Section~\ref{sec:results}. In Section~\ref{sec:discussion}, we conduct a multi-wavelength analysis using archival infrared observations, and compare the source properties to the properties of NRFs identified towards the Galactic Centre. We summarise our findings in Section~\ref{sec:conclusions}.

\section{Observations}
\label{sec:obs}

\begin{figure}
	\centering
	\includegraphics[width=\linewidth]{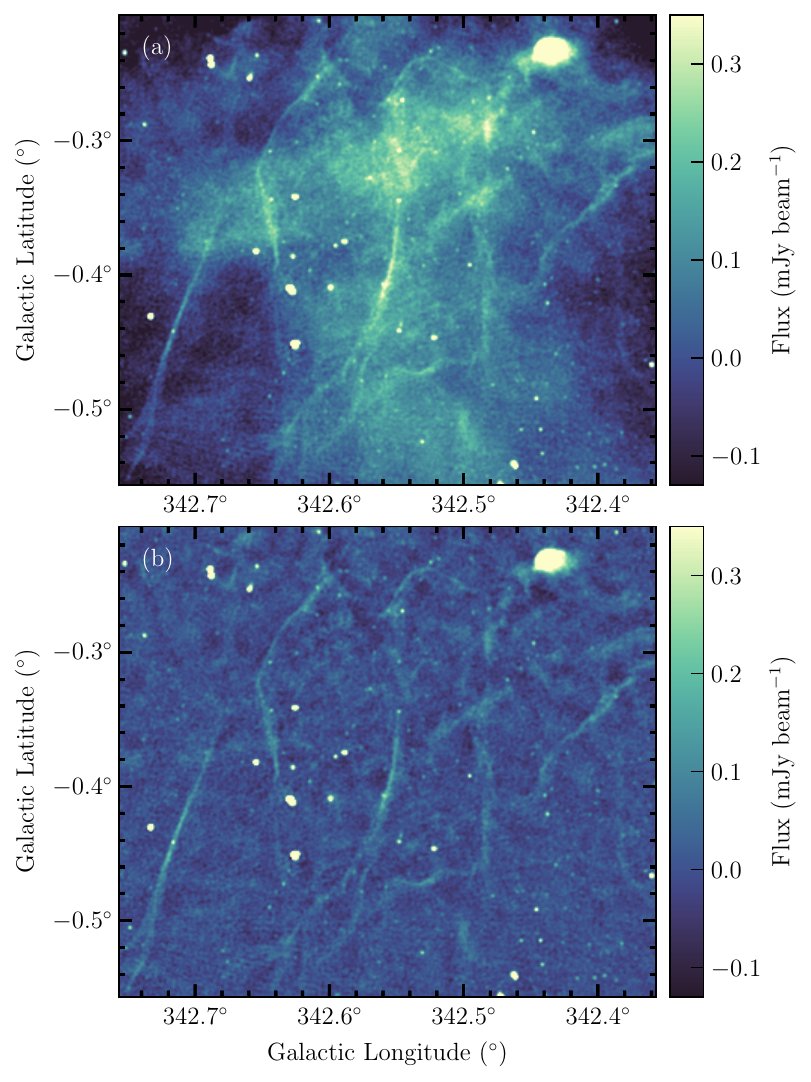} 
	\caption{(a) SMGPS 1.3\,GHz moment zero continuum towards a filamentous region in the G342.5+0.0 tile ($341^{\circ} < l < 344^{\circ}$, $-1.56^{\circ} < b < 1.56^{\circ}$). (b) The same region as shown in (a), but following the high-pass filter technique (see Section~\ref{sec:method}) for removal of diffuse emission on scales larger than 2\,arcminutes.}
	\label{fig:G342_filaments}
\end{figure}

The SARAO MeerKAT 1.3\,GHz Galactic Plane Survey \citep[SMGPS;][]{Goedhart+2024} carried out a survey of a large portion of the Galactic Plane covering Galactic longitude ranges of 2\degr\ $\le l \le$ 61\degr\ and 251\degr\ $\le l \le$ 358\degr, and latitude ranges $|b| \lesssim 1.5$\degr, using the 64-antenna MeerKAT array in the Northern Cape Province of South Africa. As detailed in \cite{Goedhart+2024}, the data were taken using the L-band receiver between 2018 Jul 21 -- 2020 Mar 14, covering the frequency range of 856 -- 1712 MHz with 4096 channels and an effective frequency of 1.3\,GHz.
Images from individual pointings were combined into 57 spatially overlapping mosaics each covering $3^{\circ}\times3^{\circ}$. We hereafter call these mosaics ``tiles'', labelled by their central coordinates in Galactic longitude and latitude. As part of the first data release (DR1), bandwidth-weighted ``zeroth moment'' integrated intensity images derived from these tiles were made available \citep[for details, see][]{Goedhart+2024}. We hereafter call these the moment zero tiles (Figure~\ref{fig:G342_filaments}(a) shows a zoomed in view of one of these moment zero tiles). The work presented throughout this paper utilises these moment zero tiles. We refer the reader to \cite{Goedhart+2024} for a comprehensive description of the data collection, reduction, calibration, imaging and DR1 data products.

\section{Method}
\label{sec:method}

\subsection{Source extraction}
\label{sec:source_extraction}

Our aim is to identify potential non-thermal radio filament (NRF) candidates in the Galactic Plane. We have therefore tuned our source extraction techniques to concentrate on filamentary structures that are reminiscent of the Galactic Centre NRF population in their properties, that is those that are bright, highly linear, and unassociated with known Galactic emission structures such as supernova remnants (SNRs). Many filament finding techniques exist, each with their own advantages and disadvantages, and each difficult in their own way to tune across a Galactic Plane survey with emission as complicated as that in the SMGPS. Our method was motivated by simplicity and computational efficiency, is semi-automated, and can be summarised by four main stages:
\begin{enumerate}
    \item Removal of large scale diffuse emission via spatial filtering
    \item Masking of emission via intensity threshold on filtered image
    \item Removal of source masks based on their morphology
    \item Manual refinement for removal of artefacts
\end{enumerate}

Large scale diffuse emission was largely removed by implementation of a spatial filtering technique \citep[e.g.][]{Yusef-Zadeh+2022A}. An initial low-pass filtered image was produced by applying a median filter to the zeroth moment tile\footnote{Using the \texttt{scipy.ndimage.median\_filter} function.}. The median filter is often used for noise reduction and edge enhancement in images, and is an appropriate approach here since our aim is to best identify the crests/ridges of filaments which may be considered similar to edges. A final high-pass filtered image was produced by subtraction of the low-pass filtered image from the original unfiltered image. The window size of the median filter was chosen by eye as that which best enhanced the filamentary features, and was equal to 15 beam widths in size ($2$\,arcminutes). The high-pass filtered images therefore have emission on scales larger than 2\,arcminutes removed. 
The effect of this filtering is illustrated in Figure~\ref{fig:G342_filaments}, which shows a filamentous region of the survey where extended emission is effectively removed. All remaining structure extraction steps were conducted on these high-pass filtered images.

Structures were identified from the high-pass filtered images via thresholding, where a mask was created of all emission above $3\times$ the rms background brightness -- we refer to these are source masks.
\cite{Goedhart+2024} found the rms background brightness to be 10-15$\mu$Jy\,beam$^{-1}$ in tiles not limited by dynamic range. After initial testing, we settled on $\sigma=20\mu$Jy\,beam$^{-1}$ for the structure extraction across all tiles, which we found to be a decent middle ground between tiles with better and worse rms noise.

By definition, the thresholding stage highlights any and all structures present in the filtered images above the set threshold, including radio galaxies, point sources, and spurious diffuse emission in higher rms noise regions for example. We thus perform morphological classification of the source masks to enable automated mask removal. Firstly, structures with an area less than 20 synthesised beam areas were removed (${\sim}2.7$\,arcminutes; a threshold judged by-eye to best remove some artefacts and small sources). The SMGPS point source catalogue (Mutale et al. in preparation) considers sources less than 5 synthesised beam areas, whilst the SMGPS extended source catalogue (Bordiu et al., under review) considers everything above 5 synthesised beam areas. All catalogues are therefore complementary and represent a combined effort to understand all of the complex emission in the SMGPS. 
Secondly, we used the $J$-plots algorithm \citep{Jaffa2018}, which from a structure's brightness distribution and shape calculates its principal moments of inertia (named $I_1$ and $I_2$) along its two principal axes. \cite{Jaffa2018} construct the $J$-moment metric as:
\begin{equation}
    J_i = \frac{I_0 - I_i}{I_0 + I_i} \, , \,\, i=1,2.
\end{equation}
where $I_0$ is the principal moment of inertia of a reference structure which is a circular disc with uniform surface brightness (which has equal principal moments of inertia along both its principal axes). In comparison to the reference value $I_0$, an elongated filament would have a smaller $I_1$ but larger $I_2$, whilst a curved filament would have both $I_1$ and $I_2$ increased. 
Therefore, for the selection of elongated filamentary structures, we select only the masks that have $J_2 < 0$ \citep[for more details, see][]{Jaffa2018}.

The final morphological criterion placed on the masks is on the aspect ratio -- defining the aspect ratio as the ratio of the mask major to minor axis, we select the masks with aspect ratio $\geq 4$ \citep[similar to the definition of a filament used in star formation studies, e.g.][]{Andre14}.\footnote{Note that the aspect ratio evaluated in this way will in the majority of cases be an underestimation of the aspect ratio of the underlying source, since the mask minor axis will be broader than the underlying structure due to the blanket inclusion of all surrounding emission $>3\sigma$.} Our final manual refinement stage was conducted via visual inspection of all extracted structures, which allowed the removal of persistent artefacts that could not be removed by our cuts. Our final catalogue contains a total of 933 elongated structures (see Figure~\ref{fig:spines}).

We stress that our filament identification process is highly conservative  in that we apply stringent cuts on morphology, size and brightness in order to identify the most isolated and elongated structures in the SMGPS images. We have identified structures that lie above an intensity threshold of $3\sigma$ (${\sim}$60$\mu$Jy\,beam$^{-1}$), are greater than 20 beam areas in size, have mask aspect ratio $>4$, and have $J_2<0$.  Therefore, our sample of 933 radio filaments is highly likely to be an underestimate of the total number of filaments in the Galactic Plane (as may be seen in comparison with the Galactic Centre in Section \ref{sec:GC}).

\begin{figure*}
	\centering
	\includegraphics[scale=0.845]{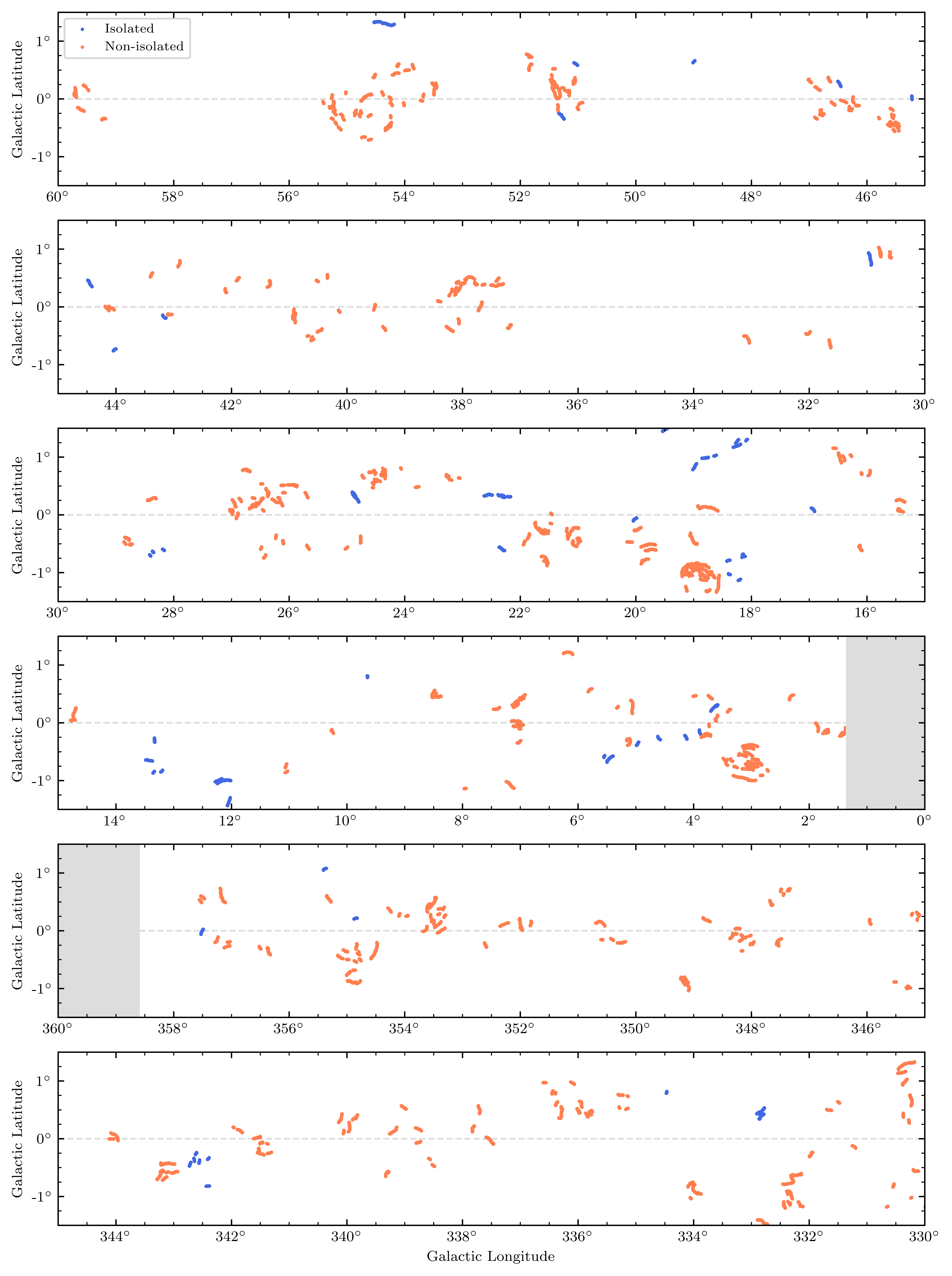}
	\caption{Positions of all identified filaments across the survey area, with their major axes marked by coloured markers (these major axes are referred to as ``spines'' from Section~\ref{sec:results} onwards). Marked in blue and orange are filaments classed as isolated and non-isolated respectively (see Section~\ref{sec:cross_corr}). Grey shaded regions indicate areas not covered by the survey (see Section~\ref{sec:obs}), and the horizontal grey dashed line marks a Galactic latitude of 0$^{\circ}$ for reference.}
	\label{fig:spines}
\end{figure*}

\begin{figure*}\ContinuedFloat
	\centering
	\includegraphics[scale=0.845]{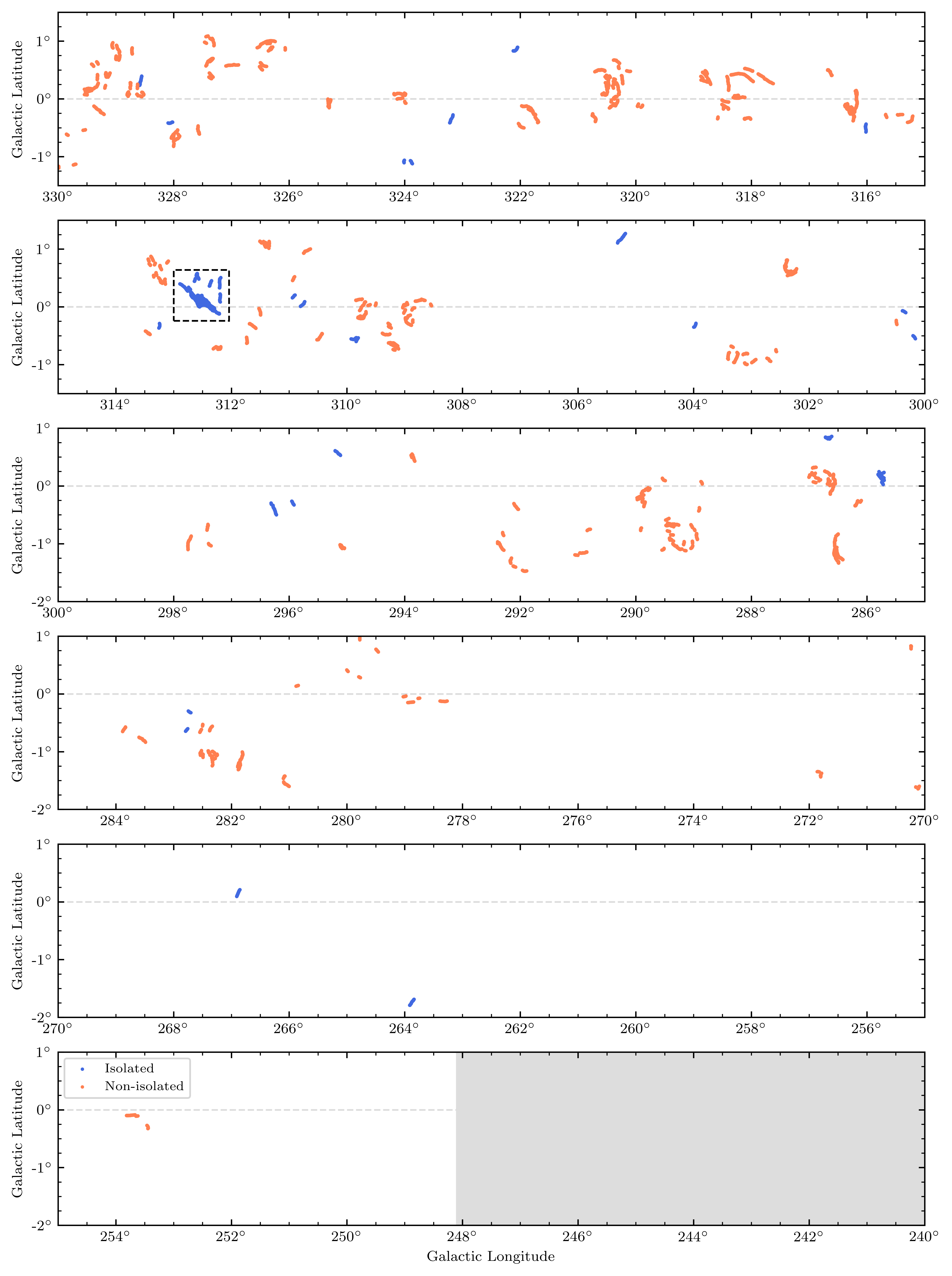}
	\caption{\textit{(cont.)} Caption as on previous page. The dashed black box marks the region presented in the SMGPS overview paper \protect\citep{Goedhart+2024}. The shift in latitude for panels with $l<300^{\circ}$ marks the shift in the SMGPS coverage to account for the Galactic warp.}
	\label{fig:spines2}
\end{figure*}

\begin{figure*}
    \centering
    \includegraphics[width=\linewidth]{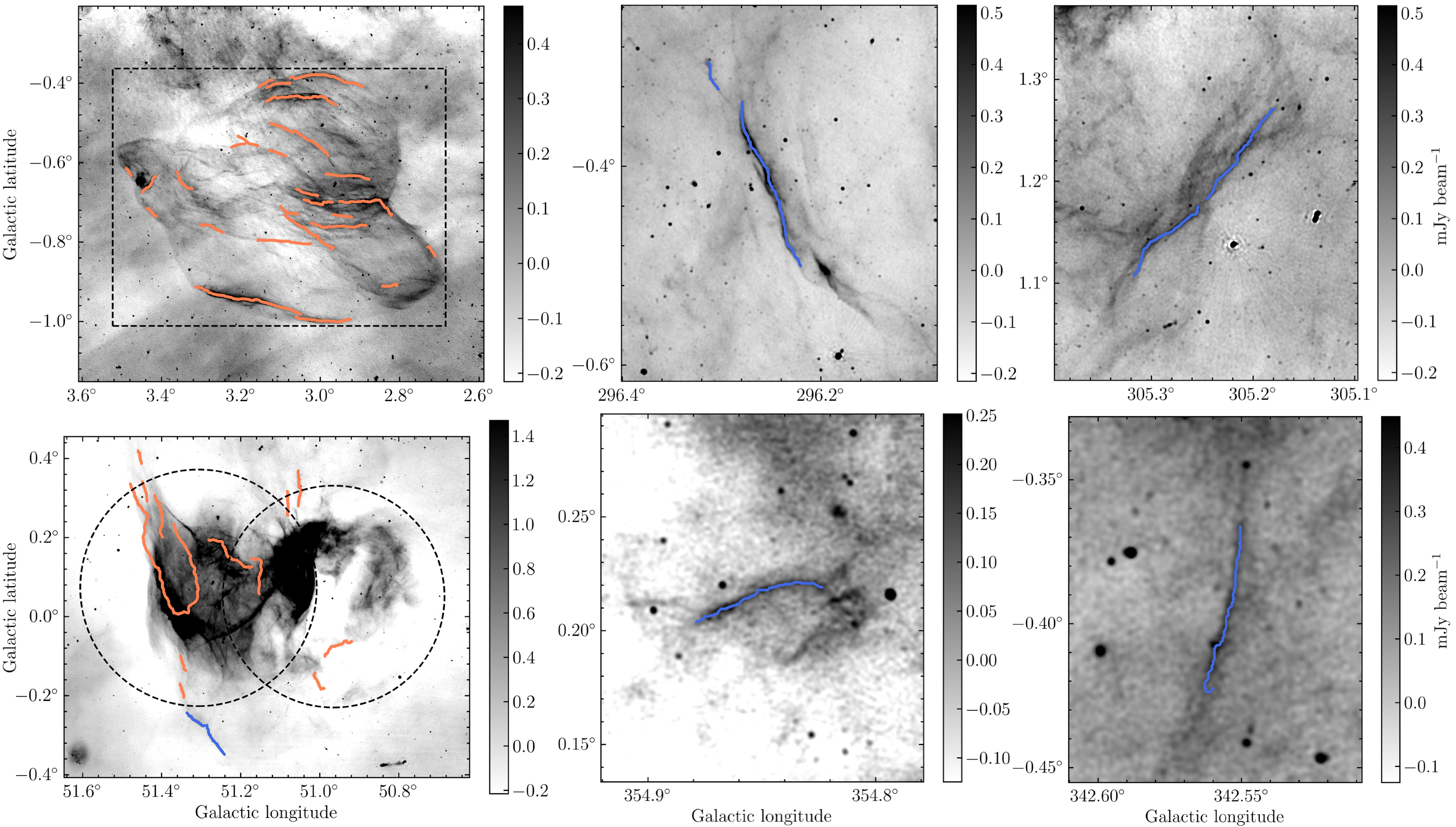}
    \caption{SMGPS 1.3\,GHz moment zero continuum (unfiltered) towards examples of identified filaments. Filaments found within extended structures are plotted in orange (i.e. non-isolated), whilst those found to be isolated are plotted in blue. Dashed black box/circles represent the bounding box or circle of known structures in the SMGPS extended source catalogue (Bordiu et al., under review).}
    \label{fig:examples}
\end{figure*}

\subsection{Cross-correlation to extended sources}
\label{sec:cross_corr}

The SMGPS extended source catalogue (Bordiu et al. under review) identified over 16,000 extended and diffuse sources, of which 24\% were matched to already known Galactic sources such as H{\sc ii} regions, supernova remnants (SNRs) and planetary nebulae.
In order to confirm that our sample of filaments are isolated structures we cross-referenced our filaments to SMGPS extended source catalogue members with radii >0.1\,degrees. This was done by-eye -- a filament was considered to be part of an extended structure if it visually overlapped with the SMGPS extended source catalogue members. We found that 86\% (803/933) of filaments overlapped with extended radio continuum structures, with the majority corresponding to SNRs \citep[also new SNR candidates in the SMGPS SNR catalogue;][]{Anderson+2024} and H{\sc ii} regions, whilst some corresponded to objects tagged by Bordiu et al. (under review) as unclassified. We refer to these as \emph{non-isolated filaments}.
The location of the remaining 14\% (130/933) of relatively isolated filaments in relation to known Galactic sources is varied. Many were found to be relatively isolated in their neighbourhood, others were found to lie just outside the catalogued radii of known Galactic sources, whilst a handful were found to reside within structures of both a relatively isolated and filamentary nature that were tagged as unclassified by Bordiu et al. (under review). 
We refer to these filaments as \emph{isolated filaments}.

\section{Results}
\label{sec:results}

We present catalogues of isolated filaments and non-isolated filaments, and make these available as part of the SMGPS DR1 (see the Data Availability statement for details, and Appendix~\ref{appendix:catalogue} for details on the catalogue entries).
The catalogues include both positional and morphological source parameters. 
For the derivation of these parameters, we reduced our filament masks to one-pixel wide ``skeletons'' using the \textsc{scikit-image} routine \textsc{skeletonize} with the "Lee" method \citep{LEE1994462}. Extraneous off-shoots (or branches) in the skeletons were pruned using the \textsc{analyze\_skeletons} function of the \textsc{FilFinder} package \citep{KochR2015}, and we hereafter refer to the pruned skeleton as the source ``spine''.\footnote{The filament spines are made available as part of the SMGPS DR1. See the Data Availability statement for details.} By definition, the filament spine traces the central crest of the filament mask. 
In this section, we describe the evaluation of our catalogued source parameters from their spines, and present our results from an analysis of these parameters across the survey area. We will present detailed analyses of individual sources in future publications.

\subsection{Source position and distribution}
\label{sec:source_pos}

We define the source positions as the geometric centroid position of a source's spine pixels. We provide these source positions in the catalogues in Galactic coordinates, pixel coordinates of the tile in which the source was identified, and in J2000 coordinates. Figure~\ref{fig:spines} shows the distribution of the source spines across the survey area, and Figure~\ref{fig:examples} shows examples of extracted filaments. The isolated filaments are coloured in blue, and are quite sparsely distributed across the survey area. Filaments that are associated with extended structure are marked in orange, and indeed many clusters of these source spines are noticeable across the survey area \citep[such as towards SNR G$3.1$-$0.6$; top-left of Figure~\ref{fig:examples};][]{Hurley-Walker+2019}. 
As seen in Figure~\ref{fig:examples}, not all the filamentary structure within SNR G$3.1$-$0.6$ is identified by our algorithm, and this is likely to be a result of the various stringent morphological cuts applied during source extraction (see \S\ref{sec:source_extraction}). 

\begin{figure}
    \centering
    \includegraphics[width=\linewidth,trim={0.24cm 0.24cm 0.24cm 0.24cm},clip]{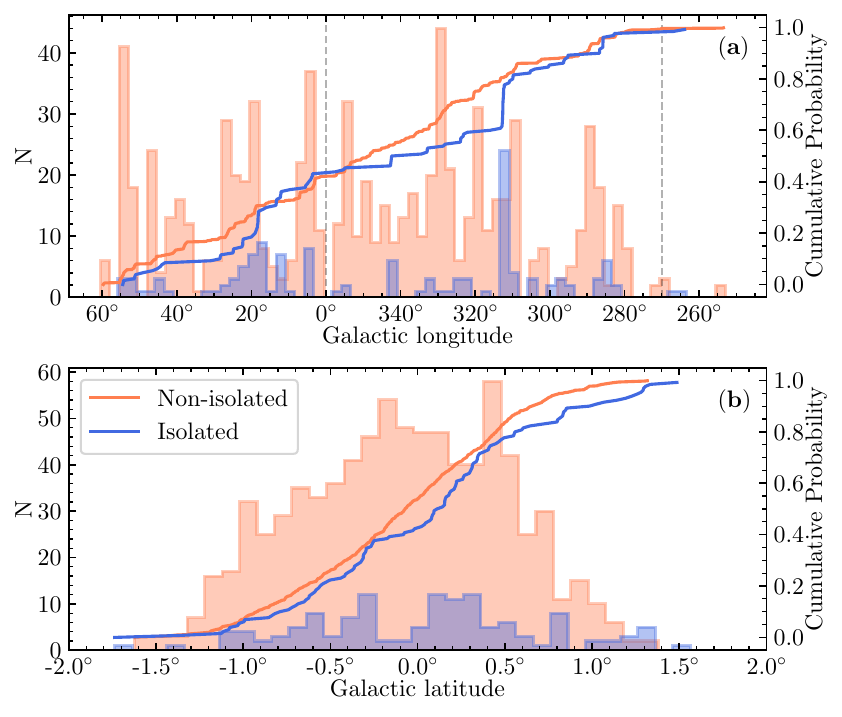} 
    \caption{Distribution of (a) Galactic longitude, and (b) Galactic latitude, of non-isolated filaments (orange), and isolated filaments (blue). The distribution is shown both as a histogram (coloured bins, left y-axis), and as a cumulative probability (solid coloured lines, right y-axis). The bin widths of the histograms are 2.5$^{\circ}$ and 0.1$^{\circ}$ in Galactic longitude and latitude respectively. Vertical grey lines in (a) delineate the three zones on the plot that correspond to the first, fourth and third Galactic quadrants (from left to right).}
    \label{fig:lon_lat_hist}
\end{figure}

Figure~\ref{fig:lon_lat_hist} shows the distributions of Galactic longitude and latitude of source centroid positions. 
The density of all filaments across Galactic longitude averages to 1.8 filaments per square degree, however it is clear from Figure~\ref{fig:lon_lat_hist} that it is not a homogeneous distribution. Filament density tails off considerably around $l<280^{\circ}$ (towards the third Galactic quadrant). This can be attributed to geometrical considerations, as our line-of-sight towards the inner Galaxy is known to pass through a higher density of spiral arms \citep[e.g.][]{DameHT2001,HouHan2014,Reid+2019}. The same is true when considering the two filament populations separately, with source densities averaging at 1.6 and 0.2\,deg$^{-2}$ across the survey area for non-isolated and isolated filaments respectively. 
Across the first and fourth Galactic quadrants, the cumulative distribution of source Galactic longitude for non-isolated filaments appears relatively uniform on the whole, whilst there appear to be two sharp increases in the isolated filament population around $l=312^{\circ}$ \citep[corresponding to the sub-sample shown in][]{Goedhart+2024} and $l=22^{\circ}$. 
Running the two-sample Kolmogorov-Smirnov (K-S) null-hypothesis test (choosing a 95\% confidence level of $p=0.05$) on the two cumulative Galactic longitude distributions returns $p=0.0004$, indicating we may reject the null hypothesis that the two distributions are drawn from the same parent distribution. 
The distribution of Galactic latitudes of non-isolated filaments appears normally distributed about $b=0^{\circ}$, whilst there are hints of a small decrease in the number of isolated filaments around $b\sim0^{\circ}$. A K-S test on the cumulative distributions of Galactic latitude returns $p=0.04$, suggesting we may only marginally reject the null hypothesis.
Having said this, it is possible that the decrease in the isolated filaments around the midplane may be a result of the difficulty in separating sources from extended Galactic emission, which is particularly bright and complex near the midplane.

\subsection{Source morphology}
\label{sec:source_morphology}

\subsubsection{Length, width, and length-to-width ratio}
\label{sec:length_width}

We define the length of the filamentary structures as the length along consecutive pixels of the one-pixel wide filament spine \citep[see also e.g.][]{ZuckerC2018}. A spine point that only varies in either its $x$ or $y$ coordinate by one compared to the previous spine point's coordinate contributes one pixel width ($\theta_{\mathrm{pix}}$) to the evaluation of the length, whilst a spine point that varies by one pixel in both its $x$ and $y$ coordinate (i.e. lies diagonally to the previous spine point) contributes $\sqrt{2}\,\times\,\theta_{\mathrm{pix}}$ to the length.

\begin{figure}
	\centering
	\includegraphics[width=\linewidth,trim={0.05cm 0cm 0.05cm 0.05cm},clip]{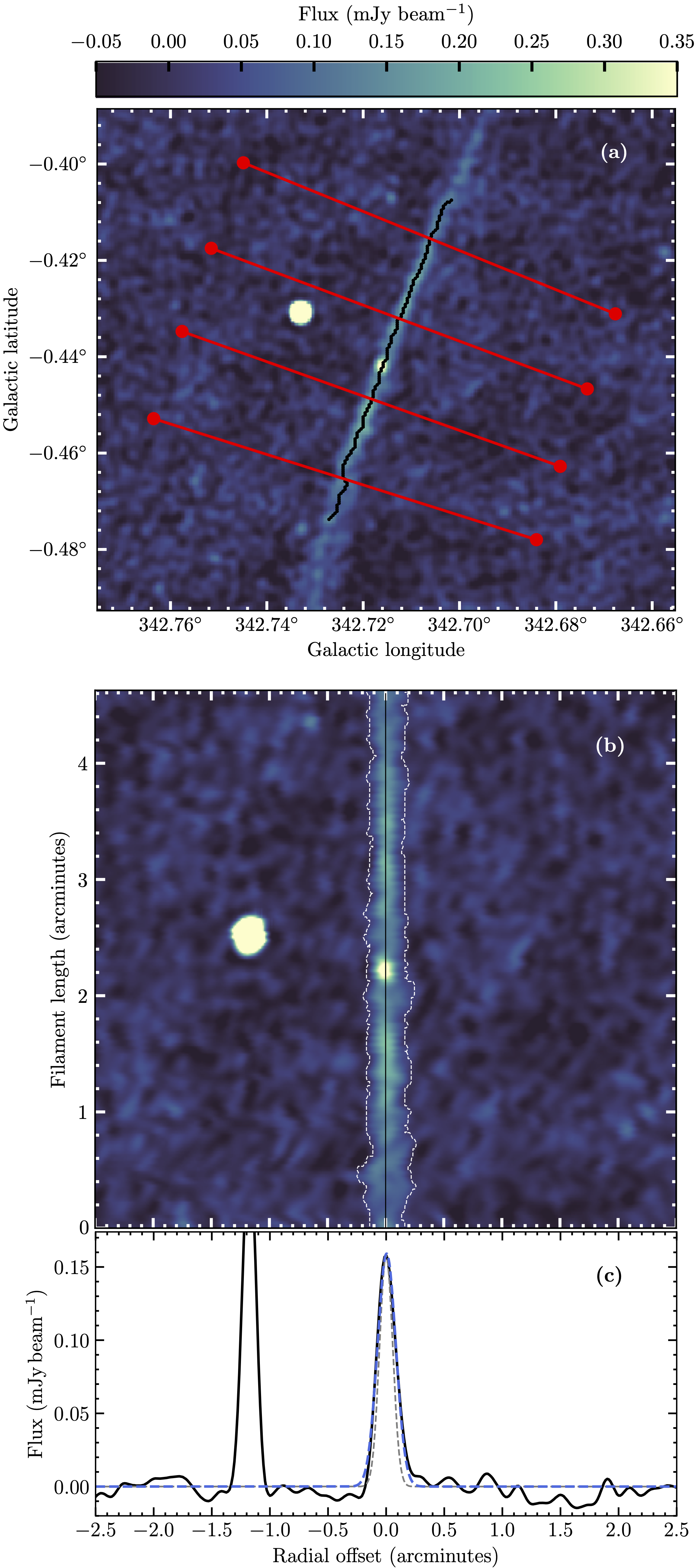} 
	\caption{(a) High-pass filtered 1.3\,GHz continuum image of an isolated filament shown in Figure~\ref{fig:G342_filaments}. The spine is marked by the black line. The red lines show four example transverse slices interpolated out to $\pm$2.5\,arcminutes in the perpendicular direction from four spine points. (b) High-pass filtered 1.3\,GHz continuum image of the filament in (a) having undergone the ``straightening'' procedure (see \S\ref{sec:length_width}). The vertical black line marks the position of the spine at a radial offset of zero, and the dashed white line marks the outline of the source mask. Panels (a) and (b) are both on the same linear colourscale. (c) The mean profile of all the transverse intensity profiles shown in panel (b). The dashed blue line is the Gaussian fit made to the profile, and the grey dashed line represents the beam shape with a FWHM of 8\,arcsec.} 
	\label{fig:transverse}
\end{figure}

We estimate the source width as the full-width at half-maximum (FWHM) of a Gaussian fitted to the mean transverse intensity profile of the source. The evaluation of this profile is however complicated by the position angle of the source on the sky (see \S\ref{sec:PA} for more on the position angle). 
We therefore implemented a straightening algorithm to effectively remove sources of their non-zero position angle. 
Using the \texttt{straighten$\_$filament$\_$interp} function of the {\sc FragMent} Python package \citep{Clarke2019,Clarke2020}, an $n^{\mathrm{th}}$ order polynomial was fitted through all the filament spine points (from testing, the best results were obtained when $n=10$).
The gradient of the polynomial at each spine point defines the normal direction of the spine (i.e. pointing along the local direction of the spine), thus the orientation of the transverse direction to the spine is defined as being perpendicular to that.
Along the transverse path to each spine point, the intensity profile was evaluated via interpolation using the second-order Taylor expansion out to $2.5$\,arcminutes either side of the spine (examples of the interpolation paths are shown in red in Figure~\ref{fig:transverse}a). In repeating this process for every spine point, and aligning each transverse intensity profile along the $x$-axis with consecutive transverse intensity profiles stacked along the $y$-axis, a ``straightened'' 2D intensity profile of the source is achieved (where the $x$-axis corresponds to the radial offset from the spine, and the $y$-axis corresponds to the filament length). The straightened source is effectively re-aligned from having a locally variable PA to having a global PA of $0^{\circ}$. The result of this procedure is shown in Figure~\ref{fig:transverse}(b), which shows the 2D transverse intensity structure of a straightened filament. 
Figure~\ref{fig:transverse}(c) shows the mean transverse intensity profile, along with the fitted Gaussian. 
We deconvolve the source width from the beam width ($\theta_{\mathrm{beam}}$) by taking $\sqrt{\theta_{\mathrm{source}}^2 - \theta_{\mathrm{beam}}^2}$. 
The filament shown in Figure~\ref{fig:transverse} has a length of 4.63\,arcminutes, PA=21\,degrees, and is resolved with a fitted FWHM of 11.2\,arcsec, and a deconvolved FWHM of 7.8\,arcseconds. 
Interestingly, this filament is one of very few in the catalogue to have a point source almost half way along its length at Galactic longitude and latitude of 342.716$^{\circ}$ and -0.442$^{\circ}$.  
This previously unidentified point source is catalogued for the first time by the SMGPS point source catalogue (Mutale et al. to be submitted), appearing under the name of G342.7160$-$0.4417 with an observed angular diameter of 18.9\,arcseconds.

\begin{figure*}
	\centering
	\includegraphics[width=\linewidth,trim={0.24cm 0.24cm 0.24cm 0.24cm},clip]{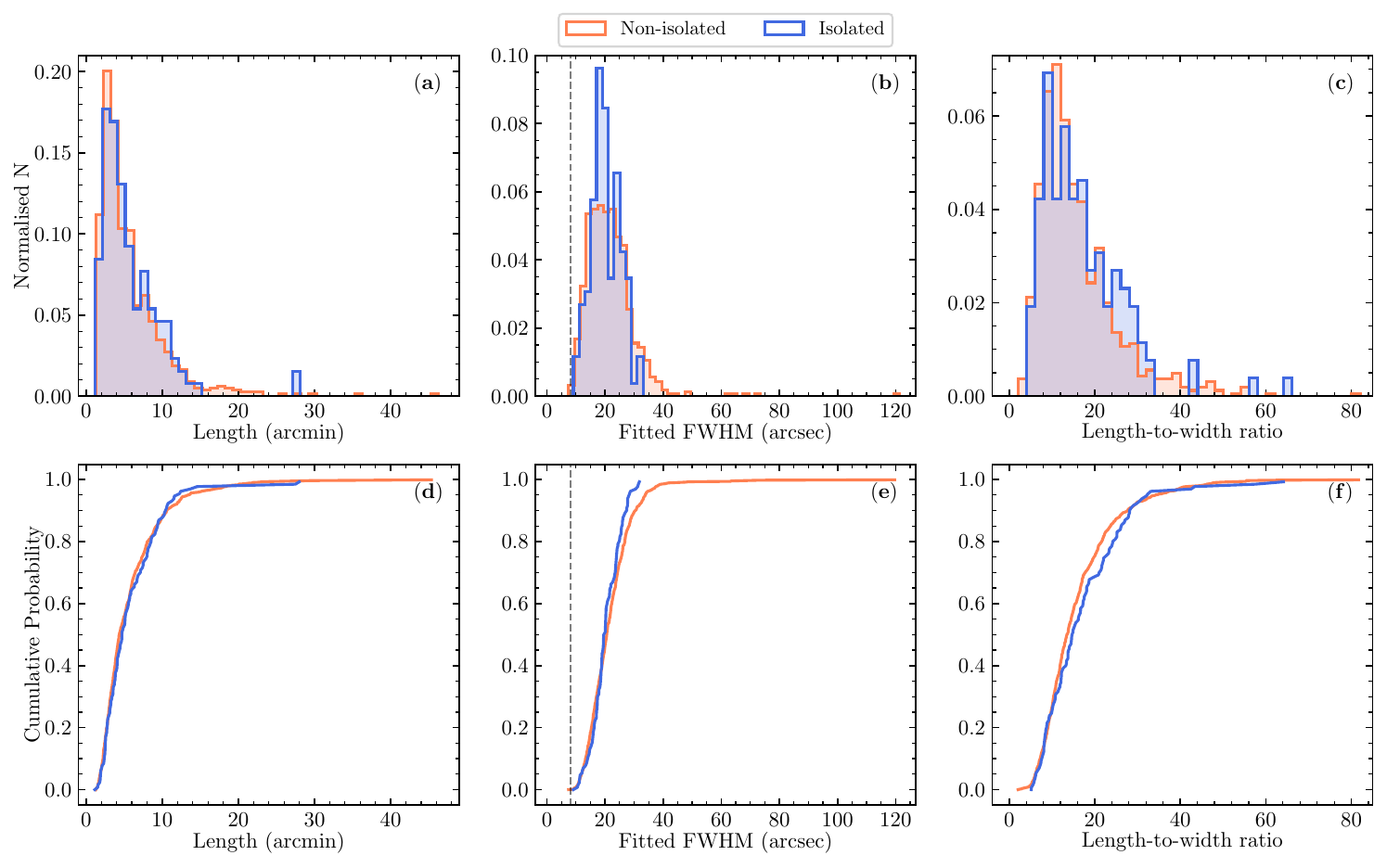}
	\caption{Normalised histograms of (a) the filament lengths in arcminutes, (b) the fitted FWHMs in arcseconds, and (c) the length-to-width ratios, of non-isolated filaments (orange) and isolated filaments (blue). The corresponding cumulative distributions are shown in panels (d), (e) and (f). In panels (b) and (e), the vertical dashed line marks the synthesised SMGPS beam width of 8\,arcsec.}
	\label{fig:length_width_aspect}
\end{figure*}

Figure~\ref{fig:length_width_aspect} shows the distributions of (a) the filament lengths, (b) the fitted FWHMs, and (c) the length-to-width ratios (defined as the ratio of the filament length to the fitted FWHM). The histograms are constructed separately for the isolated and non-isolated filaments, to allow inspection of whether the properties differ between the two source populations. 
The distribution of isolated vs non-isolated filament lengths (Figure~\ref{fig:length_width_aspect}a) by eye appear indistinguishable from each other, though a handful of non-isolated filaments are seen to extend to longer lengths. 
Running the K-S test on the cumulative distributions of filament lengths (Figure~\ref{fig:length_width_aspect}d) returns $p=0.85$, implying the two may be drawn from the same distribution.

Figure~\ref{fig:length_width_aspect}(b) shows the distribution of filament widths. At the lower end of the plot, the distributions appear similar to the eye, however it is clear that the distribution for non-isolated filaments exhibits a tail that extends to broader widths. In fact, none of the isolated filaments have widths larger than 0.57\,arcminutes. The handful of broader non-isolated filaments may be attributed to regions of extended emission being brighter and more crowded (such as the SNR region shown in Figure~\ref{fig:examples}) which contributes to more complicated transverse intensity profiles and potentially skewed fitted widths. Despite these visual difference, a K-S test run on the full range of filament widths (Figure~\ref{fig:length_width_aspect}e) returns $p=0.15$, implying the two distributions may be drawn from the same parent distribution. 
The 8\,arcsec FWHM of the MeerKAT beam is marked in Figure~\ref{fig:length_width_aspect}(b) and (e). All isolated filaments appear resolved. Only one non-isolated filament is at best marginally resolved, with a fitted FWHM of $7.5\pm0.5$\,arcsec -- this filament is seen to reside within supernova remnant G343.1$-$0.7 \citep{WhiteoakG1996}. 

During filament extraction, the aspect ratio of the source mask (defined as the mask major-to-minor axis ratio) was used as a criterion on the morphology of extracted sources (see Section~\ref{sec:source_extraction}). Here we re-calculate the aspect ratio as the ratio of the filament length to the fitted FWHM. This gives the aspect ratio of the underlying filaments within the structure masks, and we hereafter call this the length-to-width ratio. 
By definition, the length-to-width ratio of the filament will be larger than the aspect ratio of the mask due to the filament width being smaller than the mask outline. We have therefore selected the most elongated sources across the survey.
Figure~\ref{fig:length_width_aspect}(c) shows the distribution of filament length-to-width ratios. Since the distributions of filament lengths and widths each appear similar at the lower end of the distributions across the two source populations, it is unsurprising that the distributions of length-to-width ratios also appear largely similar to each other. Both peak around a median value of 14, and have standard deviations of 9.  
A K-S test on their cumulative distributions (Figure~\ref{fig:length_width_aspect}f) shows that with $p=0.28$ we may not reject the null hypothesis and the two distributions may be drawn from the same parent distribution. 
All isolated filaments have length-to-width ratios greater than 4, whilst only a handful of non-isolated filaments have length-to-width ratios less than 4. This is attributed to these non-isolated filaments being in more crowded environments resulting in poorly constrained fitted widths.

\subsubsection{Position angle}
\label{sec:PA}

The filament position angle (PA) is calculated as the arctangent of the ratio of the change in $x$ and $y$ pixel coordinates between the first and last points of the filament spine. By this definition, the position angle is measured from Galactic North (where PA = 0\,$^{\circ}$). Position angles of $\pm$90$^{\circ}$ are filaments lying parallel to the Galactic plane, where positive and negative values go towards the clockwise and anti-clockwise directions respectively.

\begin{figure*}
	\centering
	\includegraphics[width=\linewidth,trim={0.24cm 0.24cm 0.24cm 0.24cm},clip]{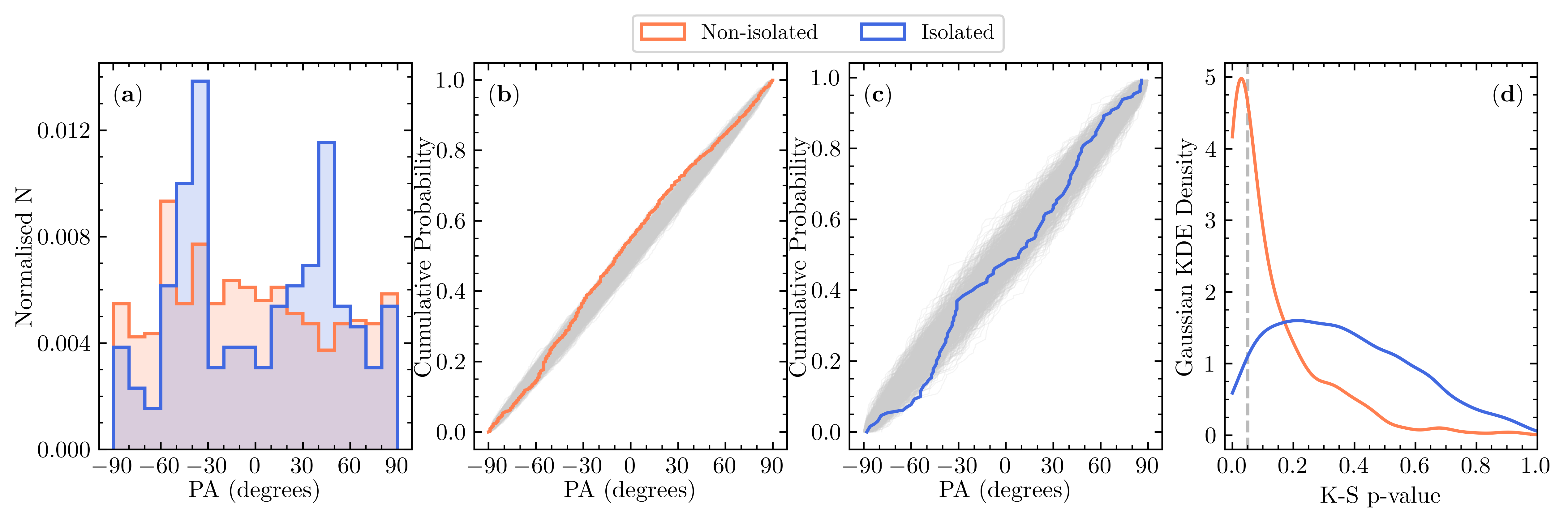} 
	\caption{(a) Normalised histogram of position angle (PA, in degrees) of non-isolated filaments in orange, and  isolated filaments in blue. (b) Cumulative distribution function (CDF) of the PAs of non-isolated filaments. Plotted in grey are 1000 realisations of randomly drawn uniform distributions between -90 and +90$^{\circ}$ (each realisation draws 803 data points i.e. the same number as the non-isolated filament sample size). (c) CDF of the PAs of isolated filaments. As in (b), grey shows the 1000 randomly drawn uniform distributions (each realisation draws 130 data points i.e. the same as the isolated filament sample size).  (d) Gaussian kernel density estimation (KDE) of the p-values resulting from a Kolmogorov-Smirnov test between each of the 1000 randomly drawn uniform distributions and the same corresponding filament sub-sample as coloured in the other panels. The vertical dashed grey line marks the 95\% confidence level at $p=0.05$.}
	\label{fig:PA_KS}
\end{figure*}

Figure~\ref{fig:PA_KS}(a) shows the normalised histograms of the PA distribution for both the isolated filaments (blue), and non-isolated filaments (orange). Both distributions extend the full range of angles, from $-$90$^{\circ}$ to +90$^{\circ}$. To the eye however, there are slight differences between the two distributions. For instance, the PA distribution for the non-isolated filaments appears relatively flat, whilst the distribution for isolated filaments appears to exhibit a slight peak in numbers around $\pm$45$^{\circ}$. This apparent difference is more clearly seen in Figures~\ref{fig:PA_KS}(b) and (c) of the cumulative distribution functions (CDFs), a visualisation of the two distributions free from any biases introduced by histogram binning. A K-S test of the two CDFs however does not corroborate these visual differences, with $p=0.16$ instead indicating that the two distributions may be drawn from the same parent distribution.

We further compare the two CDFs to that of a uniform distribution. We randomly draw 1000 uniform distributions and conduct a K-S test between each of these realisations and both PA CDFs. In each realisation, the number of randomly drawn points is matched to the filament sample size, that is we draw 803 random points for realisations compared to the non-isolated filaments, and 130 points for realisations compared to the isolated filaments. The CDFs of all randomly drawn uniform distributions are shown in grey in Figures~\ref{fig:PA_KS}(b) and (c), where it is clear there is more noise present in the random uniform distributions in Figure~\ref{fig:PA_KS}(c) where a fewer number of data points were drawn. In Figure~\ref{fig:PA_KS}(d), we show the distribution of resulting K-S test p-values, plotted using the Gaussian kernel density estimation (KDE) technique\footnote{Using the \texttt{scipy.stats.gaussian\_kde} function.}. 
The p-value distribution for the relatively isolated sources is broad, peaking around $p=0.22$, with both a high-$p$ tail and an extension down to low $p$-values below the critical value of $0.05$. Therefore in the majority of realisations, we cannot reject the null hypothesis, and the PA distribution of the isolated filament population may be said to be drawn from a uniform parent distribution. On the other hand, the non-isolated filaments peak at a p-value of 0.035, however exhibit a high-powered tail to large $p$-values. Therefore, we do not have the statistical power to say either way if the PA distribution of the non-isolated filaments is distinguishable from a uniform distribution or not.

\section{Discussion}
\label{sec:discussion}

As discussed in Section~\ref{sec:intro}, isolated non-thermal radio filaments (NRFs) have so far only been identified towards the Galactic Centre \citep{Yusef-Zadeh1984,Yusef-Zadeh2004,MeerKATCollaboration2018,Heywood2019}, and were thought to uniquely occur there due to environmental factors. 
With the SMGPS, we identify candidate isolated radio filaments outside the Galactic Centre for the first time. In this section, we cross-reference the isolated sources to infrared emission (\S\ref{sec:therm}), and compare their morphological properties to those of the Galactic Centre population as derived using our methods (\S\ref{sec:GC}).

\subsection{Thermal vs non-thermal filaments} 
\label{sec:therm}

As discussed in \cite{Goedhart+2024}, despite MeerKAT's excellent $u,v$-coverage, there is a frequency dependent minimum baseline which affects the sensitivity to bright extended emission across the band \citep[this has also been noted in other works across L band, e.g.][]{Kansabanik+2024}. 
Whilst the high-pass filtering we applied to the moment 0 images (to enhance the filamentary structures against the local background) is effective in removing the offending extended emission, the filtered images suffer significantly from negative bowl features which affect the zero-level surrounding emission. These negative bowls tend to worsen with higher frequencies, which results in an artificial steepening of the spectrum with frequency.

Due to these limitations in deriving a radio spectral energy distribution (SED) from the SMGPS data, we relied on a multiwavelength analysis to classify filaments as being thermal or non-thermal in nature. Non-thermal radio emission is typically signposted by the absence of corresponding mid-infrared emission \citep[e.g.][]{Cohen2001,Green2014,Anderson2017}, unlike thermal radio emission which is characterised by bright mid-infrared emission, typically from polycyclic aromatic hydrocarbons (PAHs). We used observations from the Midcourse Space Experiment \citep[MSX;][with 20\,arcsec angular resolution]{Price+2001} to identify \emph{candidate non-thermal radio filaments} (or candidate NRFs) that were found not to be coincident with bright mid-infrared emission upon visual inspection of the 8.3$\mu$m image.  We concentrated our analysis solely on the isolated filament population, since our cross-matching of sources to the SMGPS extended source catalogue already allowed segregation of sources known to belong to SNRs and H{\sc ii} regions. Moreover, it is the isolated filaments that have the best potential as analogues of those found in the Galactic Centre. Filaments that were spatially coincident by-eye with some bright mid-infrared emission were classed as \emph{candidate thermal radio filaments.} 
However, it is possible that the observed MIR emission are merely features that are present along the line-of-sight \citep[e.g.][]{Pare+2022,Pare+2024}. We therefore stress that without distance information in the MIR nor the radio we can only speculate on the coincidence of the MIR emission to the filaments.  In all, we identified 77/130 ($\sim$59\%) candidate non-thermal radio filaments, and 53/130 ($\sim$41\%) candidate thermal radio filaments. 
In order to firmly designate the nature of these filaments, their spectral index should be measured, but as previously stated in this section we are unable to do this with our current data. 
Given the uncertainty in the MIR classification it is possible that some candidate NRFs are misclassified as candidate thermal radio filaments -- both the candidate thermal and candidate non-thermal filaments are included in the filament catalogue detailed in Appendix~\ref{appendix:catalogue}.

Figures~\ref{fig:length_width_aspect_NT}(a)-(d) show the distribution of morphological properties for candidate thermal and non-thermal radio filaments.  
The distributions of source length, width, length-to-width ratio and position angle all appear largely similar across the two sub-populations. This is reflected by K-S tests run on their cumulative distributions (Figures~\ref{fig:length_width_aspect_NT}(e)-(h)) returning large $p$-values in all four cases ($p=0.35$, 0.11, 0.31 and 0.31 respectively for length, width, aspect ratio and PA). Therefore in the isolated filament population presented here, both the probable thermal and non-thermal filaments may have their properties drawn from the same respective parent distributions. We also examined the distributions of flux density between the candidate thermal and non-thermal filaments and found that there is no significant variation between the two sub-samples; a K-S test returns a p-value of 0.93. 

\begin{figure*}
	\centering
	\includegraphics[width=\linewidth,trim={0.24cm 0.24cm 0.24cm 0.24cm},clip]{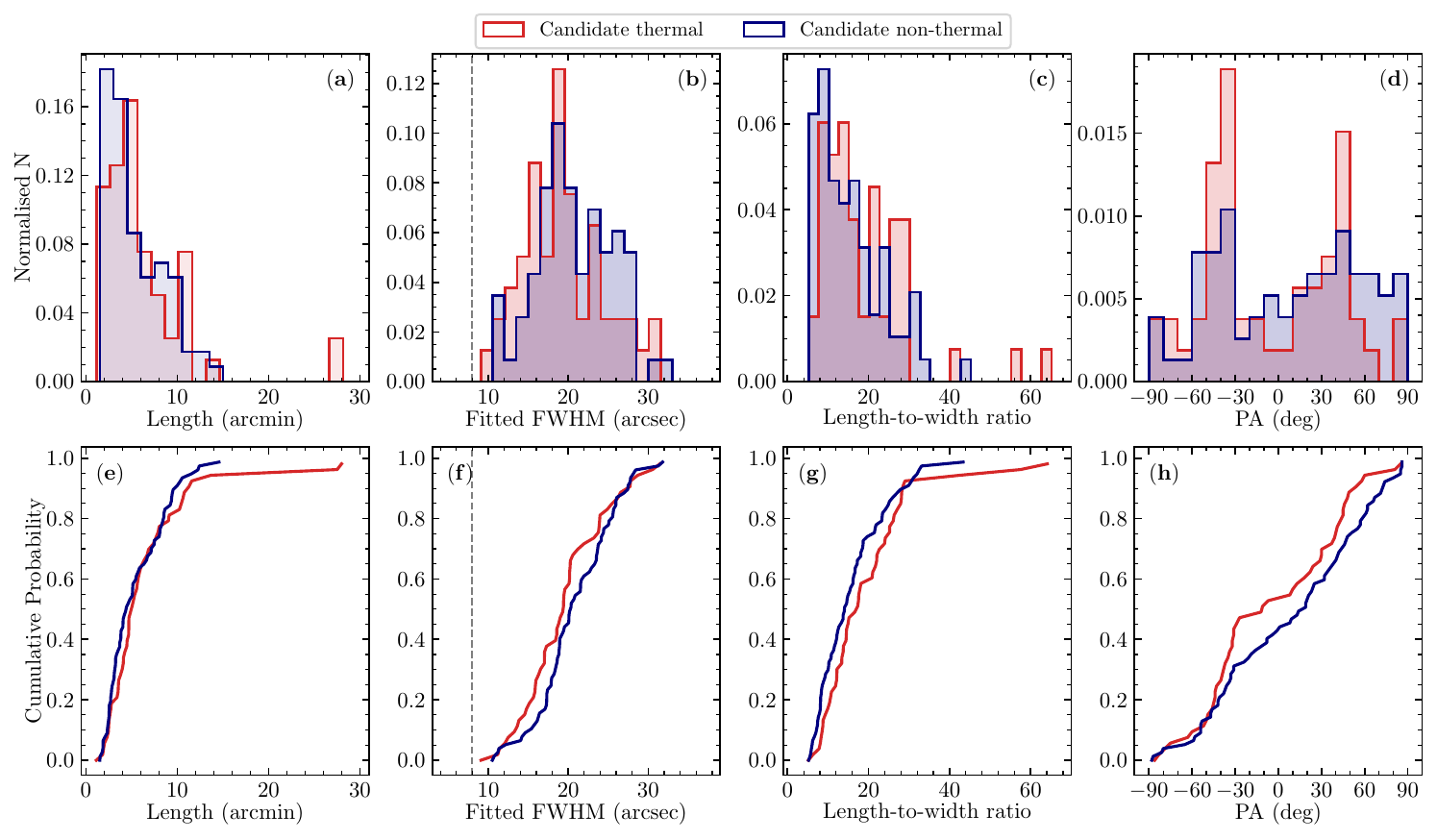}
	\caption{Normalised histograms of (a) the filament lengths in arcminutes, (b) the fitted FWHMs in arcseconds, (c) the length-to-width ratios, and (d) position angles, of isolated filaments. Those in red have coincident MSX 8.3$\mu$m infrared emission, whilst those in blue do not. Both source populations together would produce the blue histograms in Figure~\ref{fig:length_width_aspect} and \ref{fig:PA_KS}. The corresponding cumulative distributions are shown in panels (e), (f), (g) and (h). In panels (b) and (f), the vertical dashed line marks the synthesised beam width of 8\,arcsec.}
	\label{fig:length_width_aspect_NT}
\end{figure*}

\begin{figure}
	\centering
	\includegraphics[width=\linewidth,trim={0.24cm 0.24cm 0.24cm 0.24cm},clip]{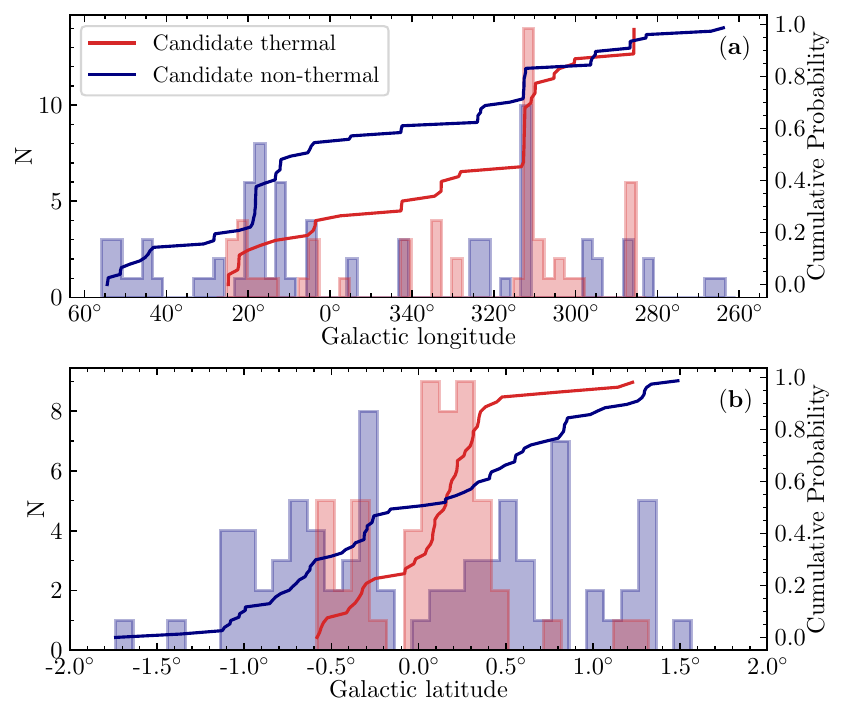}
	\caption{Distribution of (a) Galactic longitude, and (b) Galactic latitude, of the isolated SMGPS filaments expressed as a histogram (coloured bins, left y-axis), and as a cumulative distribution function (CDF; solid coloured lines, right y-axis). Red corresponds to filaments that are spatially coincident with MSX 8.3$\mu$m emission on the sky, and blue corresponds to those lacking in infrared emission on the sky. The histogram bin widths are 2.5$^{\circ}$ and 0.1$^{\circ}$ in Galactic longitude and latitude respectively.}
	\label{fig:gal_lat_hist}
\end{figure}

Figure~\ref{fig:gal_lat_hist} shows the distribution of filament centroid positions for candidate thermal and non-thermal filaments. 
Considering the Galactic longitude distribution (Figure~\ref{fig:gal_lat_hist}(a)), the longitude dependence seen in Figure~\ref{fig:lon_lat_hist} (see Section~\ref{sec:source_pos}) appears less pronounced, possibly due to the smaller sub-sample size involved.
Considering the Galactic latitude distribution (Figure~\ref{fig:gal_lat_hist}(b)), it is clear that the candidate non-thermal filaments extend farther above and below the Galactic plane, whilst the majority of those coincident with IR emission lie around the Galactic plane. This is reflected in the slope of the cumulative distribution function (CDF) between Galactic latitudes of $\pm0.5^{\circ}$ around the mid-plane being three times steeper for filaments with coincident IR emission ($0.94\pm0.05$) compared to those without ($0.31\pm0.01$). Conducting a Kolmogorov-Smirnov null hypothesis test, the returned p-value of 0.003 indicates that the null hypothesis of the two distributions being drawn from the same parent distribution can be rejected. This behaviour follows expectation, since infrared emission is known to be more concentrated toward the Galactic plane \citep[e.g.][]{MolinariS+2016} as are the massive stars and \hii\ regions which are likely to be responsible for thermally ionising radio filaments.

\subsection{Comparison to the Galactic Centre}
\label{sec:GC}

\begin{figure*}
	\centering
	\includegraphics[scale=0.88,trim={0.23cm 0.23cm 0.23cm 0.23cm},clip]{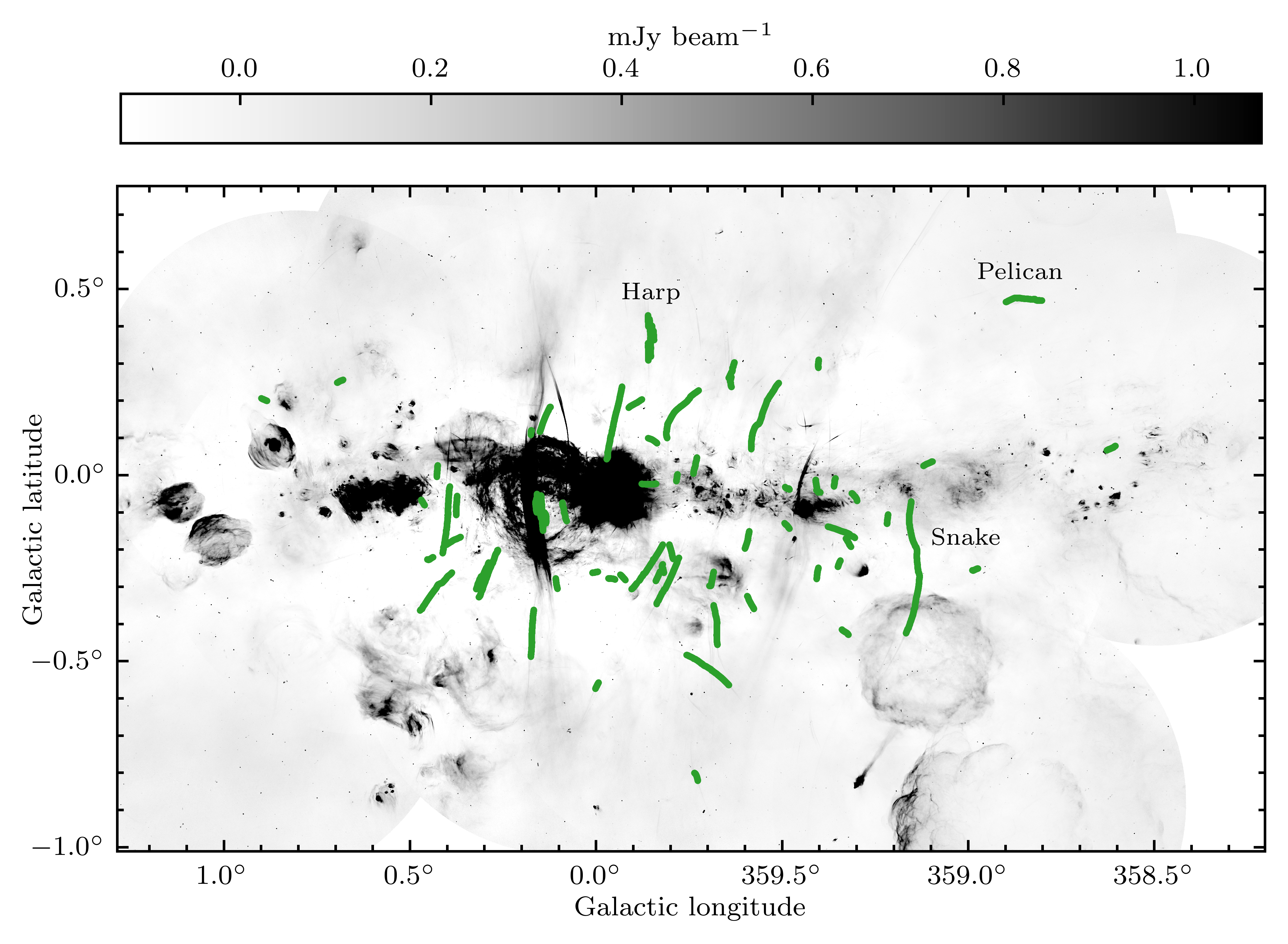}
	\caption{MeerKAT 1.28\,GHz total-intensity mosaic of the Galactic Centre \citep[][]{Heywood2022}. Overlaid in green are the one-pixel wide spines of filaments extracted using the same method as applied to the SMGPS in this work (detailed in \S\ref{sec:source_extraction}). Three named NRF regions that are mentioned in the text are labelled.}
	\label{fig:GC_spines}
\end{figure*}

As discussed in Section~\ref{sec:intro}, non-thermal radio filaments (NRFs) have until now only been seen towards the Galactic Centre \citep{Yusef-Zadeh2004,Heywood2022}. Our sample of candidate NRFs found in the SMGPS described in Section~\ref{sec:therm} are likely to contain the first NRFs to be seen outside of the Galactic Centre. Comparing the morphological properties of our candidate non-thermal filaments to the NRFs of the Galactic Centre may shed light on the origin of these filaments.

For consistency, we extracted filaments from the Galactic Centre (GC) total-intensity mosaic \citep[see Figure~\ref{fig:GC_spines};][]{Heywood2022} using exactly the same methods as applied across the SMGPS (as detailed in Section~\ref{sec:source_extraction}). At 4\,arcsec, the synthesised beam width of the GC mosaic is half that of the SMGPS\footnote{This corresponds to the synthesised beam of the Galactic Centre total-intensity mosaic being 4 times smaller in terms of beam area than the SMGPS.}. We conducted the spatial filtering of extended emission using the same 2\,arcminute size median filter that was applied to the SMGPS data (corresponding to 30 GC-mosaic beam widths).  
As with the SMGPS, we thresholded the GC mosaic to three times the rms noise, which was measured in largely emission free regions of the GC mosaic to be ${\sim}$40\,$\mu$Jy\,beam$^{-1}$. We applied the same morphological cuts on $J$-moments and mask aspect ratio as detailed in Section~\ref{sec:source_extraction}, and conducted a final by-eye visual inspection of the source masks for removal of radio galaxies and spurious sources. We also removed filaments seen to reside within known Galactic structures such as SNRs and classified the thermal/non-thermal nature of the remaining filaments using MSX 8.3\,$\mu$m emission as in Section~\ref{sec:therm}. 
As noted in Section~\ref{sec:therm} however, we caution that without distance information in the MIR nor the radio, the classification of these filaments as thermal or non-thermal may be affected by line-of-sight confusion \citep[e.g.][]{Pare+2022,Pare+2024}.
These last two steps are in addition to the procedure used by \citet{Yusef-Zadeh+2023}, but are implemented here to enable consistent cross-comparisons between GC NRFs and our SMGPS sample of candidate NRFs.

Despite the differences in our and Yusef-Zadeh's filament recovery procedures, Figure~\ref{fig:GC_spines} of our extracted GC filament spines shows we nevertheless recover the majority of the bright and long NRFs from \cite{Yusef-Zadeh+2023}. 
Significant differences arise for (i) filaments associated with bright extended emission, which are removed by our mask aspect ratio criteria, and (ii) the short and predominantly faint filament population which fail our thresholding and background criterion. 
For instance, we do not identify the majority of the Radio Arc filaments, a bundle of bright NRFs belonging to the Radio Arc bubble surrounding Sgr A* that traverse the Galactic plane through regions of extended emission. We also do not identify a significant population of the fainter filaments around the Sgr C and E regions. We do however identify well-known radio filaments such as the Snake \citep{GrayCEG1991}, Harp \citep{ThomasPE2020} and Pelican \citep{LangAKL1999}, which is encouraging as this is what our method was specifically tailored to do. In total, we identify 69 radio filaments, of which 43 lack spatially coincident infrared emission. 
In the remainder of this section, we compare the properties of the GC and candidate SMGPS NRF populations.

\subsubsection{Filament morphology}
\label{sec:GC_morph}

\begin{figure*}
	\centering
	\includegraphics[width=\linewidth,trim={0.24cm 0.24cm 0.24cm 0.24cm},clip]{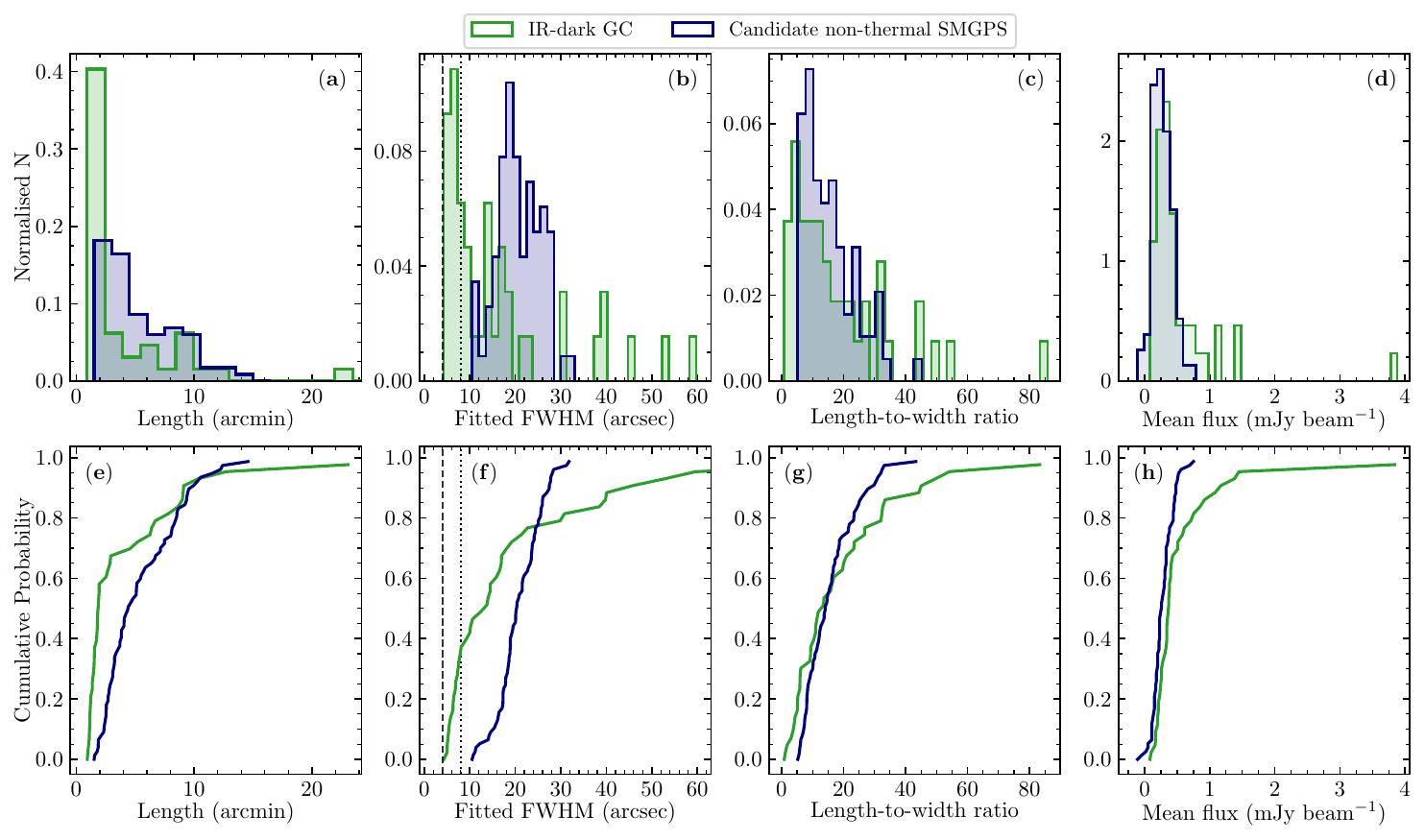} 
	\caption{Normalised histograms of (a) the filament lengths in arcminutes, (b) the fitted FWHMs in arcseconds, (c) the length-to-width ratios (i.e. the ratio of the length to the fitted FWHM), and (d) the mean flux along the source spines in mJy\,beam$^{-1}$, of the candidate non-thermal SMGPS filaments (blue) and the IR-dark Galactic Centre filaments identified using the method outlined in Section~\ref{sec:source_extraction} (green). The corresponding cumulative distributions are shown in panels (e), (f), (g) and (h). In panels (b) and (f), the vertical dotted line marks the SMGPS beam FWHM of 8\,arcsec, and the vertical dashed line marks the Galactic Centre mosaic beam FWHM of 4\,arcsec.}
	\label{fig:length_width_aspect_NT_GC}
\end{figure*}

We calculated the morphological properties of the GC filaments in the same way as detailed in Section~\ref{sec:results}, and compared them to the properties of the isolated, candidate NRFs (see Section~\ref{sec:therm}) identified across the SMGPS. 
In Figure~\ref{fig:length_width_aspect_NT_GC}, we show histograms of (a) filament lengths, (b) widths (FWHM), (c) length-to-width ratios, and (d) mean flux evaluated along the filament spines. 
We find that on the whole, the GC filaments are shorter in length than the SMGPS population, with a median length of 1.8\,pc compared to 4.4\,pc. We also find the GC filaments to be narrower in width than the SMGPS population, with a median width of 12.4\,arcseconds compared to 20.2\,arcseconds. 
A total of 5 GC filaments have widths greater than 40\,arcseconds, and these exclusively belong to those in more crowded regions such as the closely spaced Harp filaments where the transverse intensity profile fitting is confused \citep{Yusef-Zadeh+2022B}.
Running K-S tests on the cumulative distributions of source length and width (Figures~\ref{fig:length_width_aspect_NT_GC}(e) and (f)) returned vanishingly small $p$-values in both cases, indicating that these trends may not be drawn from the same parent distributions. The distributions of filament length-to-width ratios on the other hand (Figure~\ref{fig:length_width_aspect_NT_GC}(c) and (g)) appear largely similar to the eye across both the GC and the SMGPS sources, however a K-S test with a returned $p$ value of only 0.05 does not strongly corroborate these visual similarities. 

Short filaments have been identified towards the Galactic Centre in large number -- \cite{Yusef-Zadeh+2023} identify hundreds of filaments $<$66\,arcseconds in length, lying predominantly within $\pm0.4^{\circ}$ of the Galactic mid-plane. Though our method does not identify this same filament population (only 3 filaments in our extraction have lengths less than 66\,arcseconds), it is interesting that the filament population we identify indeed consists of relatively short filaments in comparison to the SMGPS. 

Either the GC filaments are indeed shorter and narrower than those identified across the SMGPS, or the difference could merely be an effect of distance in which more distant filaments appear smaller.
Many of the external mechanisms proposed for NRFs in the Galactic Centre are related to stellar winds from massive stars \citep[e.g.][]{RosnerB1996}, pulsar wind nebulae and supernova remnants \citep[e.g.][]{Bykov+2017,Barkov2019}, and other radio sources \citep[e.g.][see also Section~\ref{sec:intro}]{Yusef-Zadeh+2022C}. Assuming the same may be true of the SMGPS candidate NRFs, it is therefore reasonable to assume they may reside related to Galactic spiral arms \citep[e.g.][]{Roman-Duval+2009,Roman-Duval+2010,RigbyAJ+2019,Reid+2019}.  Were the SMGPS filament population located between us and the 8.2\,kpc distant Galactic Centre \citep{GravityCollabAAB+2019}, and assuming the intrinsic widths of the NRFs are broadly the same everywhere, the distributions of GC and SMGPS filament widths would imply that the SMGPS filaments are a factor of three closer than the GC filaments.  
We however have no constraint on where the candidate NRFs across the SMGPS reside, and therefore cannot say with any confidence that distance is responsible for the varying filament morphological properties.
It could instead be an indication of a true difference between the two populations, probably due to the more extreme properties of the Galactic Centre environment which is known to have stronger magnetic field strengths and cosmic ray ionization rates \citep[e.g.][]{Chuss+2003,vanderTak2006,Indriolo+2015}.

\subsubsection{Mean filament flux}
\label{sec:GC_flux}

We examined the distribution of mean filament flux (evaluated as the mean flux along the filament spines) of GC and SMGPS filaments in Figure \ref{fig:length_width_aspect_NT_GC}(d). As can be seen, the GC filaments have a tail to higher fluxes than compared to the SMGPS filaments. A K-S on their cumulative distributions (Figure~\ref{fig:length_width_aspect_NT_GC}(h)) returns $p=0.005$, meaning we may reject the null hypothesis and say that the two distributions may not be drawn from the same parent distribution.
If the differences in length and width between the GC and SMGPS populations are due to the greater distances of the GC filaments this implies that the GC filaments are much more luminous than the SMGPS population. It is difficult to derive the total energy of synchrotron radiation without underlying knowledge of the local magnetic field strength \citep[which is notoriously unconstrained in the GC;][]{Crocker2010}. Nevertheless, assuming equipartition, an intrinsically more luminous GC filament population would imply greater cosmic ray electron energies, consistent with the order of magnitude increase in cosmic ray ionization rates seen in the GC versus local molecular clouds \citep[e.g.][]{vanderTak2006,Indriolo+2015}.

\subsubsection{Position angle}
\label{sec:GC_PA}

\begin{figure*}
	\centering
	\includegraphics[width=\linewidth,trim={0.24cm 0.24cm 0.24cm 0.24cm},clip]{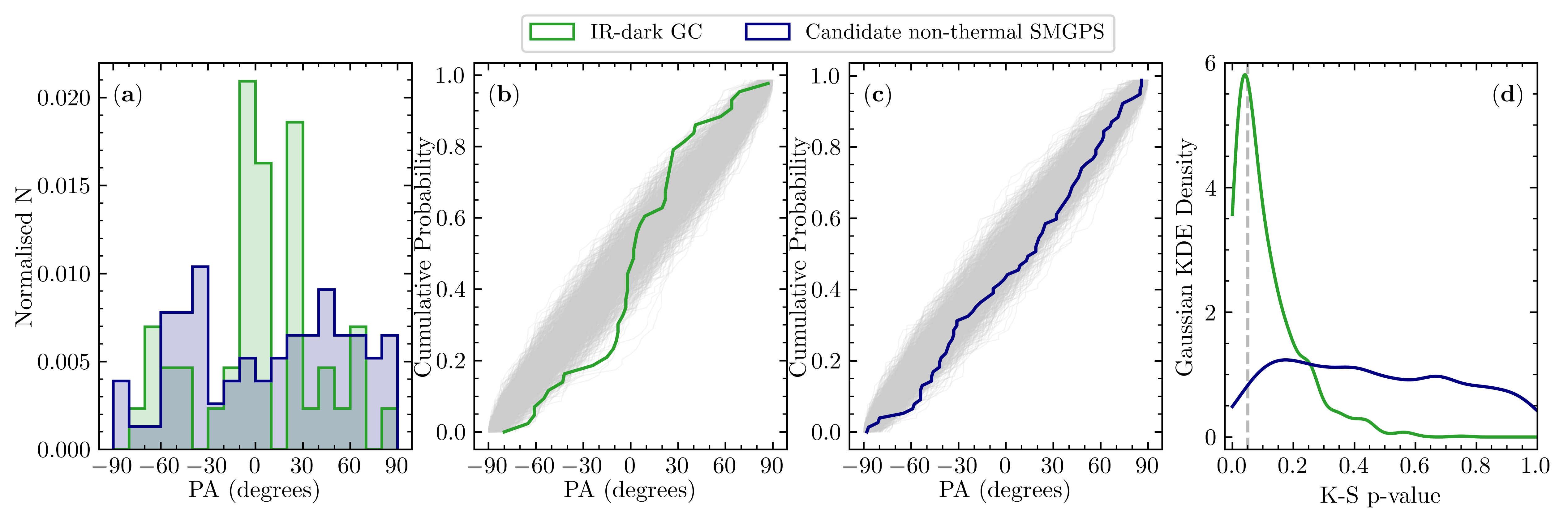} 
	\caption{(a) Normalised histogram of position angle (PA, in degrees) of of isolated, candidate non-thermal SMGPS sources in blue, and IR-dark GC filaments in green. (b) Cumulative distribution function (CDF) of the PAs of IR-dark GC filaments. Plotted in grey are 1000 realisations of randomly drawn uniform distributions between -90 and +90$^{\circ}$ (each realisation draws 69 data points i.e. the same number as the IR-dark GC sample size). (c) CDF of the PAs of candidate non-thermal SMGPS filaments. As in (b), grey shows the 1000 randomly drawn uniform distributions (each realisation draws 77 data points i.e. the same as the candidate non-thermal SMGPS filament sample size).  (d) Gaussian kernel density estimation (KDE) of the p-values resulting from a Kolmogorov-Smirnov test between each of the 1000 randomly drawn uniform distributions and the same corresponding filament sub-sample as coloured in the other panels. The vertical dashed grey line marks the 95\% confidence level at $p=0.05$.}
	\label{fig:PA_NT_GC}
\end{figure*}

Figure~\ref{fig:PA_NT_GC}(a) shows the distribution of filament position angles (PA) for both the IR-dark GC and candidate non-thermal SMGPS populations. The PAs of the SMGPS filaments was shown in Section~\ref{sec:PA} to likely be drawn from a uniform distribution. The PAs of the IR-dark GC filaments on the other hand appear clearly peaked around PA=0$^{\circ}$. However, as in Section~\ref{sec:PA}, a K-S test on the two cumulative distributions (Figure~\ref{fig:PA_NT_GC}(b) and (c)) does not corroborate these visual differences, with $p=0.13$ instead indicating that the two distributions may be drawn from the same parent distribution. We repeated the analysis presented in Section~\ref{sec:PA}, and compared the distributions of PAs to 1000 randomly drawn uniform distributions (see Figure~\ref{fig:PA_NT_GC}b). As in Section~\ref{sec:PA}, the number of data points that were randomly drawn were matched to the filament sample size, that is we drew 69 random points for realisations compared to the IR-dark GC sample, and 77 points for realisations compared to the candidate non-thermal SMGPS sample. These random uniform realisations are shown in Figures~\ref{fig:PA_NT_GC}(b) and (c).
A K-S test was conducted between each of the 1000 random realisations and their corresponding filament sample, and the resulting $p$-values are plotted in Figure~\ref{fig:PA_NT_GC}(d). The distribution of $p$-values of the candidate non-thermal SMGPS filaments sit firmly above the critical $p$-value of 0.05, meaning we the two distributions may be drawn from the same parent distribution (i.e. that of a uniform distribution). Though the distribution of $p$-values of the IR-dark GC filaments peaks at 0.04, ${\sim}$70\% of the 1000 K-S tests returned a $p$-value $>$0.05, with a median $p$-value above 0.05 of 0.13.

\begin{figure}
	\centering
	\includegraphics[width=\linewidth,trim={0.23cm 0.23cm 0.23cm 0.23cm},clip]{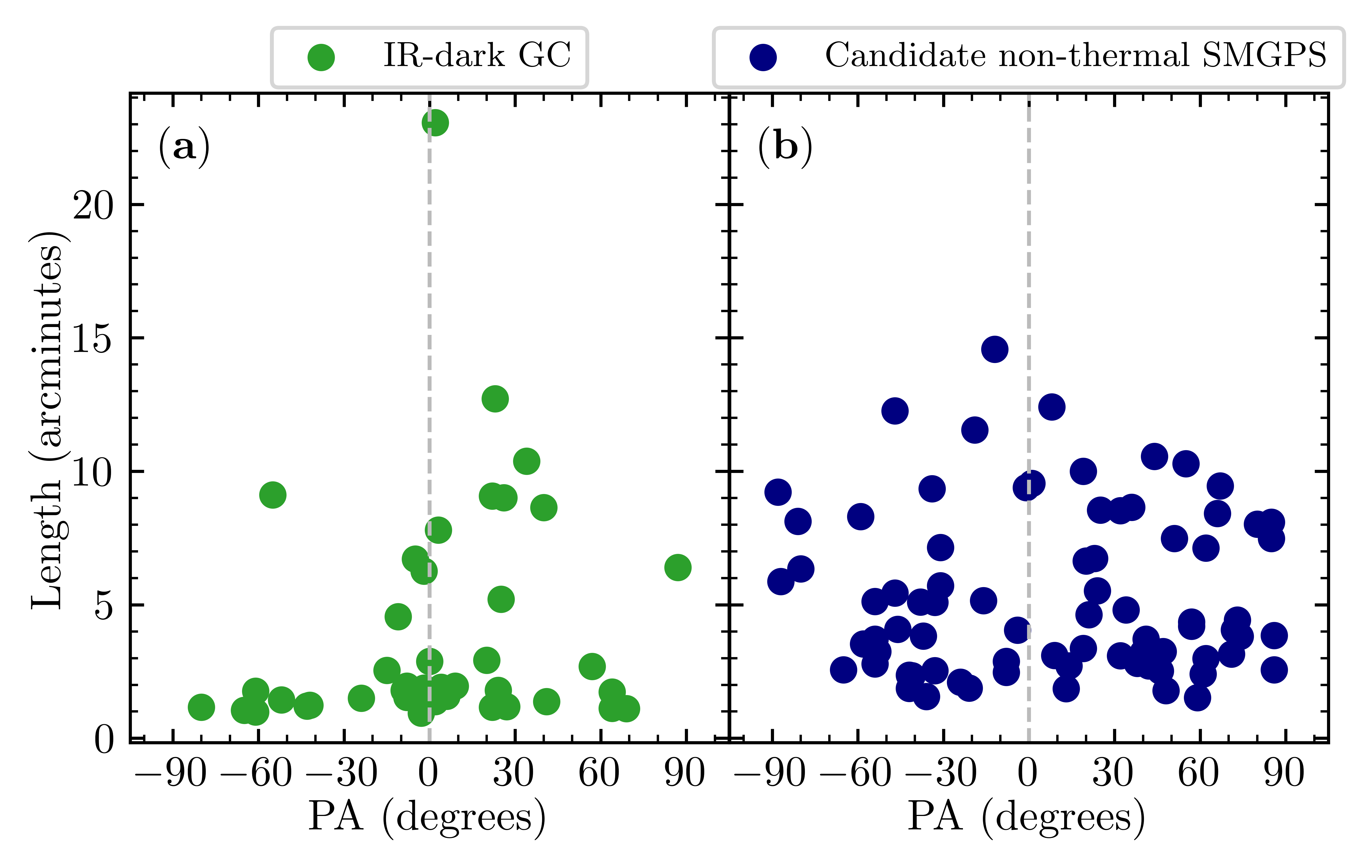}
	\caption{Distribution of filament position angle (PA) against length for (a) the IR-dark Galactic Centre filaments, and (b) the IR-dark SMGPS filaments (see \S\ref{sec:therm}). The vertical grey dashed line marks PA$=0^{\circ}$.}
	\label{fig:PA_vs_length}
\end{figure}

\cite{Yusef-Zadeh+2023} present an analysis of the position angles of GC filaments with respect to their lengths. They show that the short filament population (with lengths $<$1\,arcminute) have a largely uniform distribution of position angles with respect to Galactic North, whilst the longest filaments exhibit position angles largely perpendicular to the Galactic Plane. As shown in Figure~\ref{fig:PA_vs_length}(a), we also recover this behaviour, despite the smaller sample of filaments that our method was able to extract. However, the SMGPS candidate NRFs do not show the same relation between the PA and their length (see Figure~\ref{fig:PA_vs_length}(b)).

The Milky Way has an ordered large scale magnetic field in the form of an axi-symmetric or bi-symmetric spiral, which is similar to that of other nearby galaxies \citep{han2002}. Filaments tracing this large scale magnetic field should be expected to be centred around position angles of $\pm90$\degr, i.e.~parallel to the Galactic Plane. The magnetic field towards the Galactic Centre is known to be bimodal and stronger than that of molecular clouds in the Galactic disc \citep[e.g.][]{Chuss+2003,MorrisM2006,ferriere2009}. Whilst infrared and sub-mm studies of the GC have shown the presence of a large scale magnetic field oriented parallel to the Galactic Plane \citep[e.g.][]{Nishiyama+2010,Mangilli+2019,Guan+2021,Butterfield+2024,Pare+2024}, radio studies have shown the GC NRFs that are oriented perpendicularly to the Galactic Plane trace an ordered, vertical magnetic field \citep[e.g.][]{Tsuboi+1986,Yusef-ZadehWP+1997,LangME1999}.  Given the SMGPS filaments have an apparently uniform distribution of position angles from $-90$ to $+90$\,degrees, we suggest that the local magnetic field the SMGPS filaments trace does not follow the large scale Galactic field.

\subsubsection{Filament Galactic distribution}

\begin{figure}
    \centering
    \includegraphics[width=\linewidth,trim={0.24cm 0.24cm 0.24cm 0.24cm},clip]{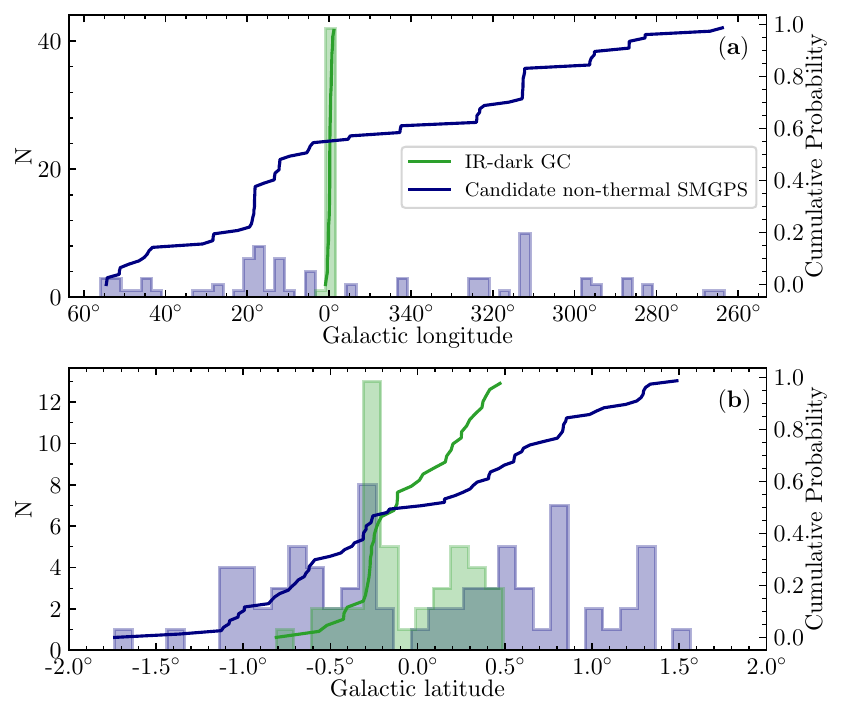}
    \caption{Distribution of (a) Galactic longitude, and (b) Galactic latitude, of the candidate non-thermal SMGPS filaments (blue), and the IR-dark GC filaments (green). The distribution is shown both as a histogram (coloured bins, left y-axis), and as a cumulative probability (solid coloured lines, right y-axis). The bin widths of the histograms are 2.5$^{\circ}$ and 0.1$^{\circ}$ in Galactic longitude and latitude respectively.}
    \label{fig:lon_lat_hist_GC}
\end{figure}

Finally we compared the angular distribution of the Galactic Centre (GC) and SMGPS candidate NRFs to each other. Figure \ref{fig:lon_lat_hist_GC} shows histograms of the GC and SMGPS candidate NRFs in both Galactic longitude and latitude. Immediately one can see in the upper panel that the Galactic Centre has a much higher density of filaments than the rest of the Galactic Plane. Non-thermal radio filaments may not be unique to the GC, but the unusual properties of the GC certainly have resulted in a much greater density of filaments. 

The lower panel of Figure \ref{fig:lon_lat_hist_GC} shows the Galactic latitude distribution of both filament populations. The GC filaments tend to lie closer to $b=0^{\circ}$ than their SMGPS counterparts, although there is a deficit at $b=0^{\circ}$ in both populations. A K-S test suggests it is unlikely that both samples are drawn from the same population (with $p=0.002$). The deficit in both populations at $b=0^{\circ}$ may be due to source confusion and the difficulty in identifying filaments in the complex emission near the mid-plane. However, the significant difference in latitude distribution between the two populations is likely to be a real effect as both samples were extracted with the same techniques.  \citet{Heywood2022} speculate that the GC filaments are causally connected to a large scale radio bubble associated with Sgr A* \citep{Heywood2019}. We suggest that the greater density of filaments toward $b=0^{\circ}$ in the GC provides supporting evidence for this hypothesis.

\subsubsection{Correlation of SMGPS vs GC filament morphological parameters regardless of mid-infrared classification} 
\label{sec:ignore_MIR}

\begin{figure*}
	\centering
	\includegraphics[width=\linewidth,trim={0.24cm 0.24cm 0.24cm 0.24cm},clip]{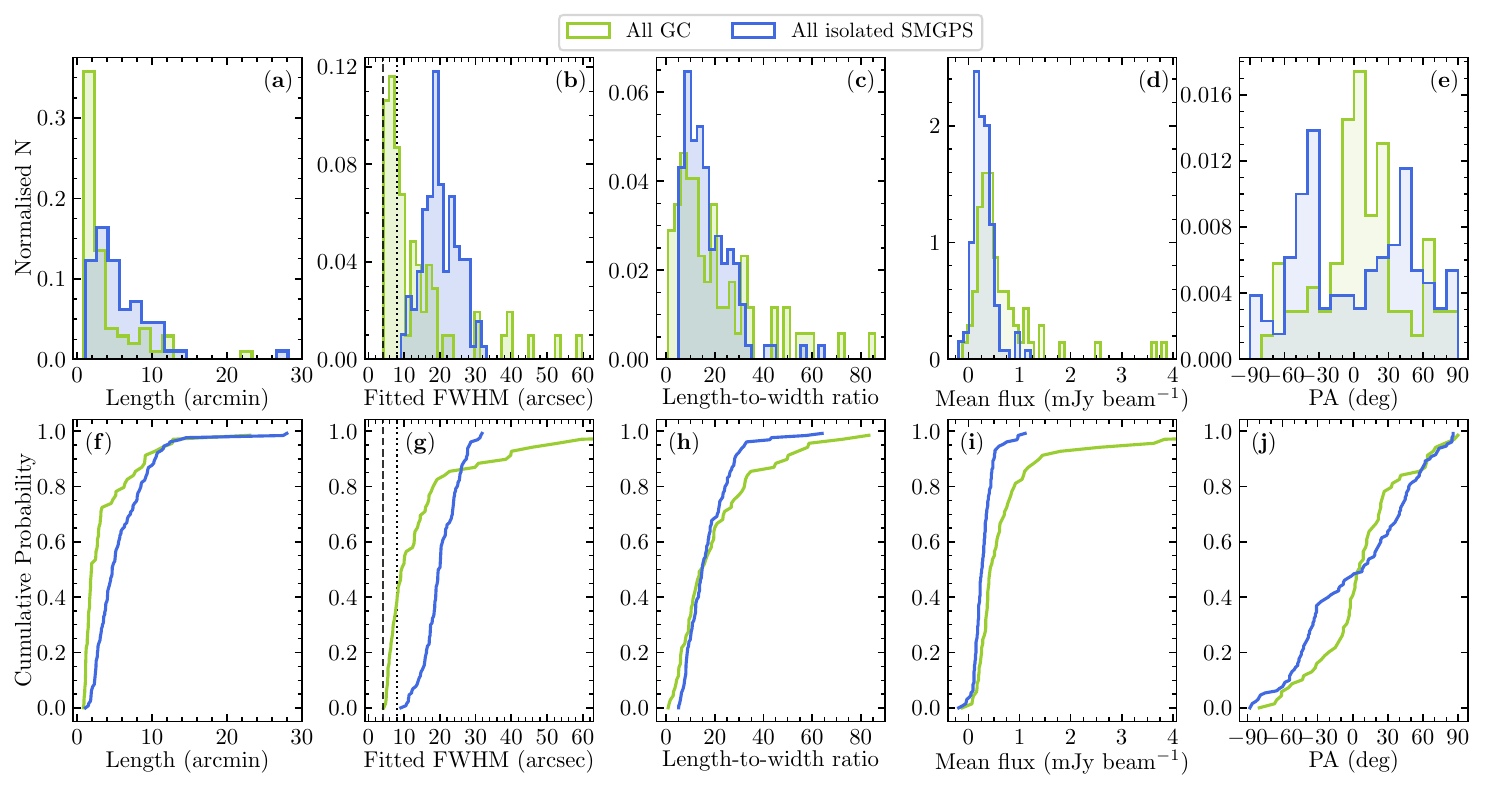} 
	\caption{Normalised histograms of (a) the filament lengths in arcminutes, (b) the fitted FWHMs in arcseconds, (c) the length-to-width ratios (i.e. the ratio of the length to the fitted FWHM), (d) the mean flux along the source spines in mJy\,beam$^{-1}$, and (e) position angles, of all isolated SMGPS filaments (blue, i.e. the same distributions shown in Figures~\ref{fig:length_width_aspect} and \ref{fig:PA_KS}) and all Galactic Centre filaments identified using the method outlined in Sections~\ref{sec:source_extraction} and \ref{sec:GC} (green; i.e. without considering the mid-infrared analysis). The corresponding cumulative distributions are shown in panels (f), (g), (h), (i) and (j). In panels (b) and (g), the vertical dotted line marks the SMGPS beam FWHM of 8\,arcsec, and the vertical dashed line marks the Galactic Centre mosaic beam FWHM of 4\,arcsec.}
	\label{fig:length_width_aspect_ALL_GC}
\end{figure*}

In Sections~\ref{sec:therm} and \ref{sec:GC}, we classified our SMGPS and Galactic Centre filament populations as candidate thermal and candidate non-thermal based on the by-eye coincidence (or lack thereof) of 8.3\,$\mu$m mid-infrared (MIR) emission observed with MSX. Widespread MIR emission of both the GC and (inner) Galactic plane may result in a large number of filaments being misidentified as candidate thermal filaments 
due to line-of-sight confusion \citep[e.g.][]{Pare+2022,Pare+2024}. We therefore present a comparison of the distributions of morphological parameters for the full \textit{isolated} filament populations extracted across the SMGPS (in Section~\ref{sec:method}) and towards the GC (in Section~\ref{sec:GC}) regardless of the MIR classification (see Figure~\ref{fig:length_width_aspect_ALL_GC}).

The correlations discussed in Sections~\ref{sec:GC_morph} and \ref{sec:GC_flux} are unchanged. We find the filaments, regardless of our MIR classification, remain both shorter and narrower towards the GC than across the SMGPS (Figure~\ref{fig:length_width_aspect_ALL_GC}(a), (b), (f) and (g)). We continue to find the distribution of length-to-width ratios similar across both the GC and SMGPS samples (Figure~\ref{fig:length_width_aspect_ALL_GC}(c) and (h)), and that the distribution of mean fluxes extends to larger values in the GC population than across the SMGPS (Figure~\ref{fig:length_width_aspect_ALL_GC}(d) and (i)). K-S tests run on the cumulative distributions of filament lengths, widths, length-to-width ratios and mean fluxes all corroborate the correlations found towards the candidate non-thermal filaments in Sections~\ref{sec:GC_morph} and \ref{sec:GC_flux}. In fact, the returned $p$-values indicate a strengthening of the noted correlations (see Table~\ref{tab:pvalues}), likely attributable to the increased sample size of filaments considered compared to the candidate non-thermal only filaments shown in Figure~\ref{fig:length_width_aspect_NT_GC}. 

In Section~\ref{sec:GC_PA}, the position angle distribution of the candidate non-thermal filaments across the SMGPS and the IR-dark GC filaments were visually different (Figure~\ref{fig:PA_NT_GC}(a)), however this was not strongly corroborated by a K-S test which returned a $p$-value of 0.13. An analysis of the two distributions in relation to randomly drawn uniform distributions however revealed that whilst the SMGPS population may be drawn from a uniform distribution that the GC distribution could not. Now, regardless of the MIR classification (Figure~\ref{fig:length_width_aspect_ALL_GC}(e) and (j)), we find that the K-S run on the cumulative distributions of isolated SMGPS and GC filaments returns a smaller $p$-value of 0.03, corroborating those visual differences and indicating they may not be drawn from the same parent distribution. This change in $p$-value is also likely attributable to the increase in sample size.

In sum, we conclude that our discussion of the properties of the SMGPS filaments in relation to the GC filaments is robust to the uncertain MIR classification.

\begin{table}
    \centering
    \caption{$p$-values returned by running Kolmogorov-Smirnov (K-S) tests on the cumulative distributions of the morphological parameters of SMGPS filaments and Galactic Centre filaments. Column 2 pertains to tests run on only the candidate non-thermal isolated filament populations (i.e. the distributions shown in Figures~\ref{fig:length_width_aspect_NT_GC} and \ref{fig:PA_NT_GC}), whilst column 3 shows the tests run on all the isolated filaments (including both candidate non-thermal and candidate thermal filaments; i.e. the distributions shown in Figure~\ref{fig:length_width_aspect_ALL_GC}).  }  
    \label{tab:pvalues}
    \setlength\tabcolsep{4.5pt}
    \begin{tabular}{lll}
    \hline\hline
    Parameter               & Candidate NRFs only                    & All isolated filaments  \\
                            & (\S\ref{sec:GC_morph}, \ref{sec:GC_flux} and \ref{sec:GC_PA})                                             & (\S\ref{sec:ignore_MIR}) \\ \hline
    Length                  & $1.4\times10^{-7}$                          & $1.9\times10^{-9}$ \\
    Width                   & $1.2\times10^{-7}$                          & $2.0\times10^{-15}$  \\ 
    Length-to-width ratio   & 0.05                                        & 0.13 \\
    Mean flux               & 0.005                                       & $1.5\times10^{-6}$ \\ 
    PA                      & 0.13                                            & 0.03 \\ \hline
    \end{tabular}
\end{table}

\section{Conclusions}
\label{sec:conclusions}

Here we present a catalogue of 933 filamentary structures observed in the 1.3\,GHz continuum SARAO MeerKAT Galactic Plane Survey (SMGPS). Our filament identification process is tuned to identify a conservative catalogue of the brightest, most extended and unconfused filaments in the images and includes a final visual refinement to remove artefacts.

We divide our sample into 803 non-isolated filaments that are associated with extended radio sources in the SMGPS Extended Source Catalogue (Bordiu et al., under review) and 130 isolated filaments. Non-isolated filaments are predominantly associated with supernova remnants (SNRs) and \hii\ regions. A multiwavelength analysis reveals that 77 of the isolated filaments are excellent candidates for non-thermal radio filaments like those seen toward the Galactic Centre \citep{Yusef-Zadeh+2023,Heywood2022}. If so, these are the \emph{first} NRFs observed outside the Galactic Centre (GC), calling into question theories of their formation that rely on the unique properties within the GC.

Using the same extraction process as we used for the SMGPS, we extracted a consistent comparison sample of non-thermal filaments from the MeerKAT Galactic Centre mosaic published by  \cite{Heywood2019}. We compare the two populations of filaments in detail to ascertain their similarities and difference. In summary we find:

\begin{itemize}

\item The GC filaments are found to be predominantly shorter and narrower than the SMGPS population, with many GC filaments only marginally resolved by the synthesised beam. The morphologies of the two populations are different to each other, with K-S tests indicating it is statistically unlikely that the two samples are drawn from the same parent distribution. We speculate that this could be an effect of distance were the GC filaments 
more distant. 
However without accurate distances to the SMGPS filaments we cannot be certain.   

\item The mean flux densities of GC filaments have a brighter tail to the distribution, although the bulk of both populations have similar mean flux densities. We suggest that the GC filaments are intrinsically more luminous than the SMGPS population, implying a more energetic cosmic ray population in the GC, however in the absence of distance information we cannot conclusively say.

\item We find no evidence for the Position Angle (PA) distribution of the longest SMGPS filaments being preferentially aligned perpendicular to the Galactic Plane, in contrast to that found by \cite{Yusef-Zadeh+2023}. We confirm Yusef-Zadeh's result for the GC with our independent GC filament extraction. Considering the overall large scale magnetic field structure of the Milky Way, it is probable that the local field traced by the SMGPS filament population does not follow the global magnetic field.

\item Lastly, we compare the overall Galactic distribution of filaments and show that the GC has a much greater density of known filaments than the Galactic Plane. Although non-thermal radio filaments are not unique to the GC, the unusual properties within the GC have certainly resulted in a greater number of filaments. We detect a significant difference in the latitude distribution of GC and SMGPS filaments, which supports the causal connection with Sgr A* already suggested by \cite{Heywood2022}.

\end{itemize}

In summary we have identified a population of candidate non-thermal radio filaments (NRFs) within the SARAO MeerKAT Galactic Plane Survey (SMGPS), the first such population identified outside the Galactic Centre. Whilst a detailed analysis of the filament radio spectral indices was not possible in this work, these candidate NRFs warrant future studies to confirm their nature and associate them to individual cosmic ray emitters. We will present studies of individual filaments in future publications.

\section*{Acknowledgements}

We thank the referee, Dylan Par\'e, for a constructive report that helped improve the quality of this paper. 
GMW and MAT gratefully acknowledge the support of the UK's Science \& Technology Facilities Council (STFC) through grant awards ST/R000905/1 and ST/W00125X/1. 
The MeerKAT telescope is operated by the South African Radio Astronomy Observatory (SARAO), which is a facility of the National Research Foundation, an agency of the Department of Science and Innovation.
This research has made use of NASA's Astrophysics Data System Bibliographic Services, and {\sc python} packages {\sc astropy} \citep{astropy}, {\sc cmocean} \citep{cmocean}, FragMent \citep{Clarke2019}, $J$-plots \citep{Jaffa2018}, {\sc matplotlib} \citep{matplotlib}, {\sc numpy} \citep{numpy}, {\sc pandas} \citep{pandas2010}, RadFil \citep{ZuckerC2018}, FilFinder \citep{KochR2015}, {\sc scipy} \citep{scipy} and {\sc scikit-image} \citep{scikit-image}.

\section*{Data Availability}

The filament catalogues and the filament spine masks are made available at: \url{https://doi.org/10.48479/9k9c-ey22}.
The SMGPS Data Release 1 \citep[DR1;][]{Goedhart+2024} containing the 1.3\,GHz images analysed here are available at: \url{https://doi.org/10.48479/3wfd-e270}.


\bibliographystyle{mnras}
\bibliography{example} 



\appendix

\section{Catalogues}
\label{appendix:catalogue}

We present two filament catalogues, one of filaments identified to reside within mostly known Galactic sources (non-isolated filaments), and a second of comparatively isolated filaments (see Section~\ref{sec:cross_corr}). The isolated filament catalogue contains both the candidate thermal and candidate non-thermal filaments classified by our mid-infrared analysis (see Section~\ref{sec:therm}). Both catalogues follow the structure detailed in Table~\ref{tab:filaments}.

\begin{table*}
    \centering
    \caption{Format of the catalogue of filamentary structures.}    
    \label{tab:filaments}
    \begin{tabular}{lll}
    \hline\hline
    Column name     & Unit                      & Description \\ \hline
    ID              & --                        & Filament ID number, unique within the isolated catalogue, and unique within the non-isolated catalogue.\\
    tile            & --                        & The first four characters of the name of the image tile. \\
    lon             & deg                       & Galactic longitude of the centroid position of the source skeleton.  \\ 
    lat             & deg                       & Galactic latitude of the centroid position of the source skeleton. \\
    x\_coord        & pix                       & x pixel coordinate within the tile of the centroid position of the source skeleton. \\
    y\_coord        & pix               & y pixel coordinate within the tile of the centroid position of the source skeleton. \\
    RA              & $^{\mathrm{h\,m\,s}}$ [J2000] & Right Ascension of the centroid position of the source skeleton. \\
    Dec             & $^{\circ}$ $'$ $''$ [J2000] & Declination of the centroid position of the source skeleton. \\
    bbox\_x1         & pix               & x pixel coordinate of the bottom left corner of the source bounding box. \\
    bbox\_y1         & pix               & y pixel coordinate of the bottom left corner of the source bounding box. \\
    bbox\_x2         & pix               & x pixel coordinate of the top right corner of the source bounding box. \\
    bbox\_y2         & pix               & y pixel coordinate of the top right corner of the source bounding box. \\
    peak\_flux       & mJy beam$^{-1}$   & Peak flux of the source along the skeleton pixels only. \\
    mean\_flux       & mJy beam$^{-1}$   & Mean flux of the source along the skeleton pixels only. \\
    length          & arcmin            & Length of the source skeleton. \\
    width\_fwhm      & arcsec            & Fitted FWHM of the source derived from Gaussian fitting.\\
    width\_deconv    & arcsec            & Width of the source deconvolved from the 8\,arcsec MeerKAT beam.\\ 
    width\_err & arcsec           & Statistical error on the FWHM from the Gaussian fitting. \\
    aspect        & --                & Aspect ratio of source mask, derived during source extraction as the ratio of the mask major and minor axes evaluated \\
                    &                   & using the \textsc{python} function \texttt{skimage.measure.regionprops}. \\
    l\_to\_w\_ratio        & --                & Length-to-width ratio of the source, derived from the calculated length and fitted width (columns length$/$width\_fwhm). \\
    angle           & deg               & Orientation of the source skeleton with respect to Galactic North (i.e 0\,degrees is perpendicular to the Galactic Plane, \\
                    &                   & positive values tend clockwise and negative values tend anti-clockwise). \\
    J1              & --                & $J_1$ moment \\
    J2              & --                & $J_2$ moment \\ 
    assoc\_source   & --                & The name of the associated extended sources (if any) as it appears in the SMGPS extended source catalogue \\ 
                        &                   & (Bordiu et al., under review). \\

    MSX\_emission    & --                & Only for the isolated filament catalogue, a flag denoting whether the source is coincident with infrared 8.3$\mu$m MSX\\
                            &                   & emission. \\ \hline
    \end{tabular}
\end{table*}


\bsp	
\label{lastpage}
\end{document}